\documentclass[longauth]{aa}
\usepackage{graphicx}
\usepackage{epsf,color}
\usepackage[mathscr]{eucal}
\usepackage{amsmath}
\usepackage{amssymb,amsfonts}
\usepackage{natbib}
\usepackage[varg]{txfonts}
\usepackage{dsfont}
\usepackage[modulo]{lineno}
\definecolor{Mygreen}{rgb}{0.00, 0.72, 0.0}
\definecolor{Mypink}{rgb}{1.0, 0.0, 0.5}
\usepackage{float}
\usepackage{color} 
\usepackage{cuted}
\usepackage{appendix}
\usepackage{multirow}

\definecolor{pastelorange}{rgb}{1.0, 0.7, 0.28}

\bibpunct{(}{)}{;}{a}{}{,}

\usepackage[citecolor=blue, linkcolor=Mygreen, urlcolor=Mypink, colorlinks=true]{hyperref}

\begin{document}

\title{Continuum, CO, and water vapour maps of the Orion Nebula}
\subtitle{First millimetre spectral imaging with CONCERTO}
\date{Received  28 April 2025 / Accepted 28 July 2025}

\author{
    F.-X.~D\'esert \inst{\ref{IPAG}}
    \and J.~F.~Mac\'ias-P\'erez \inst{\ref{LPSC}}
    \and  A.~Beelen \inst{\ref{LAM}}
     \and  A.~Beno\^it \inst{\ref{Neel}}
     \and  M.~Béthermin \inst{\ref{stg}}
     \and  J.~Bounmy \inst{\ref{LPSC}}
     \and  O.~Bourrion \inst{\ref{LPSC}}
     \and  M.~Calvo \inst{\ref{Neel}}
     \and  A.~Catalano \inst{\ref{LPSC}}
     \and  C.~De Breuck\inst{\ref{eso_germany}}
     \and C.~Dubois\inst{\ref{LAM}}
     \and C.~A~Dur$\rm\acute{a}$n\inst{\ref{IRAME}}
     \and  A.~Fasano \inst{\ref{LAM}, \ref{iac}, \ref{ull}}
     \and  J.~Goupy \inst{\ref{Neel}}
     \and  W.~Hu \inst{\ref{LAM}, \ref{uwc}}
     \and  E.~Ibar \inst{\ref{Valparaiso}, \ref{Mingal}}
     \and  G.~Lagache \inst{\ref{LAM}}
     \and  A.~Lundgren \inst{\ref{LAM}}
     \and  A.~Monfardini \inst{\ref{Neel}}
     \and  N.~Ponthieu \inst{\ref{IPAG}}
     \and  D.~Quinatoa \inst{\ref{Valparaiso}}
     \and  M. Van Cuyck \inst{\ref{LAM}}
     \and  R.~Adam \inst{\ref{OCA}}
     \and  P.~Ade \inst{\ref{Cardiff}}
     \and  H.~Ajeddig \inst{\ref{CEA}}
     \and  S.~Amarantidis \inst{\ref{IRAME}}
     \and  P.~Andr\'e \inst{\ref{CEA}}
     \and  H.~Aussel \inst{\ref{CEA}}
     \and  S.~Berta \inst{\ref{IRAMF}}
     \and  A.~Bongiovanni \inst{\ref{IRAME}}
     \and  D.~Ch\'erouvrier \inst{\ref{LPSC}}
     \and  M.~De~Petris \inst{\ref{Roma}}
     \and  S.~Doyle \inst{\ref{Cardiff}}
     \and  E.~F.~C.~Driessen \inst{\ref{IRAMF}}
     \and  G.~Ejlali \inst{\ref{Teheran}}
     \and  A.~Ferragamo \inst{\ref{Roma}}
     \and  A.~Gomez \inst{\ref{CAB}} 
     \and  C.~Hanser \inst{\ref{LPSC}}
     \and  S.~Katsioli \inst{\ref{AthenObs}, \ref{AthenUniv}}
     \and  F.~K\'eruzor\'e \inst{\ref{Argonne}}
     \and  C.~Kramer \inst{\ref{IRAMF}}
     \and  B.~Ladjelate \inst{\ref{IRAME}} 
     \and  S.~Leclercq \inst{\ref{IRAMF}}
     \and  J.-F.~Lestrade \inst{\ref{LERMA}}
     \and  S.~C.~Madden \inst{\ref{CEA}}
     \and  A.~Maury \inst{\ref{Barcelona1}, \ref{Barcelona2}}
     \and  F.~Mayet \inst{\ref{LPSC}}
     \and  A.~Moyer-Anin \inst{\ref{LPSC}}
     \and  M.~Mu\~noz-Echeverr\'ia \inst{\ref{IRAP}}
     \and  I.~Myserlis \inst{\ref{IRAME}}
     \and  A.~Paliwal \inst{\ref{Roma2}}
     \and  L.~Perotto \inst{\ref{LPSC}}
     \and  G.~Pisano \inst{\ref{Roma}}
     \and  V.~Rev\'eret \inst{\ref{CEA}}
     \and  A.~J.~Rigby \inst{\ref{Leeds}}
     \and  A.~Ritacco \inst{\ref{LPSC}}
     \and  H.~Roussel \inst{\ref{IAP}}
     \and  F.~Ruppin \inst{\ref{IP2I}}
     \and  M.~S\'anchez-Portal \inst{\ref{IRAME}}
     \and  S.~Savorgnano \inst{\ref{LPSC}}
     \and  A.~Sievers \inst{\ref{IRAME}}
     \and  C.~Tucker \inst{\ref{Cardiff}}
     \and  R.~Zylka \inst{\ref{IRAMF}}
     }

   \institute{
     Univ. Grenoble Alpes, CNRS, IPAG, 38000 Grenoble, France 
     \label{IPAG}
     \and 
      Univ. Grenoble Alpes, CNRS, Grenoble INP, LPSC-IN2P3, 53, avenue des Martyrs, 38000 Grenoble, France
     \label{LPSC}
     \and
     Aix Marseille Univ, CNRS, CNES, LAM (Laboratoire d'Astrophysique de Marseille), Marseille, France
     \label{LAM}
     \and
     Institut N\'eel, CNRS, Universit\'e Grenoble Alpes, France
     \label{Neel}
     \and
     Université de Strasbourg, CNRS, Observatoire astronomique de Strasbourg, UMR 7550, 67000 Strasbourg, France
     \label{stg}
     \and
     European Southern Observatory, Karl Schwarzschild Straße 2, 85748 Garching, Germany
     \label{eso_germany}
     \and
     Institut de Radioastronomie Millim\'etrique (IRAM), Avenida Divina Pastora 7, Local 20, E-18012, Granada, Spain
     \label{IRAME}     
     \and       
     Institut de RadioAstronomie Millim\'etrique (IRAM), Grenoble, France
     \label{IRAMF}
     \and
     Instituto de F\'isica y Astronom\'ia, Universidad de Valpara\'iso, Avda. Gran Breta\~na 1111, Valpara\'iso, Chile
     \label{Valparaiso}
     \and
     Millennium Nucleus for Galaxies (MINGAL)
     \label{Mingal}
     \and
     Universit\'e C\^ote d'Azur, Observatoire de la C\^ote d'Azur, CNRS, Laboratoire Lagrange, France 
     \label{OCA}
     \and
     School of Physics and Astronomy, Cardiff University, Queen’s Buildings, The Parade, Cardiff, CF24 3AA, UK 
     \label{Cardiff}
     \and
     Universit\'e Paris-Saclay, Universit\'e Paris Cit\'e, CEA, CNRS, AIM, 91191, Gif-sur-Yvette, France
     \label{CEA}
     \and
     Dipartimento di Fisica, Sapienza Universit\`a di Roma, Piazzale Aldo Moro 5, I-00185 Roma, Italy
     \label{Roma}
     \and
     Institute for Research in Fundamental Sciences (IPM), School of Astronomy, Tehran, Iran
     \label{Teheran}
     \and
     Centro de Astrobiolog\'ia (CSIC-INTA), Torrej\'on de Ardoz, 28850 Madrid, Spain
     \label{CAB}
     \and
     Instituto de Astrofísica de Canarias, E-38205 La Laguna, Tenerife, Spain
     \label{iac}
     \and
     Departamento de Astrofísica, Universidad de La Laguna (ULL), E-38206 La Laguna, Tenerife, Spain
     \label{ull}
     \and
     Department of Physics and Astronomy, University of the Western Cape, Robert Sobukhwe Road, Bellville, 7535, South Africa
     \label{uwc}
     \and    
     National Observatory of Athens, Institute for Astronomy, Astrophysics, Space Applications and Remote Sensing, Ioannou Metaxa
     and Vasileos Pavlou GR-15236, Athens, Greece
     \label{AthenObs}
     \and
     Department of Astrophysics, Astronomy \& Mechanics, Faculty of Physics, University of Athens, Panepistimiopolis, GR-15784
     Zografos, Athens, Greece
     \label{AthenUniv}
     \and
     High Energy Physics Division, Argonne National Laboratory, 9700 South Cass Avenue, Lemont, IL 60439, USA
     \label{Argonne}
     \and  
     LERMA, Observatoire de Paris, PSL Research University, CNRS, Sorbonne Universit\'e, UPMC, 75014 Paris, France  
     \label{LERMA}
     \and
     Institute of Space Sciences (ICE), CSIC, Campus UAB, Carrer de Can Magrans s/n, E-08193, Barcelona, Spain
     \label{Barcelona1}
     \and
     ICREA, Pg. Lluís Companys 23, Barcelona, Spain
     \label{Barcelona2}
     \and
     IRAP, CNRS, Université de Toulouse, CNES, UT3-UPS, (Toulouse), France 
     \label{IRAP}
     \and
     Dipartimento di Fisica, Universit\`a di Roma ‘Tor Vergata’, Via della Ricerca Scientifica 1, I-00133 Roma, Italy        
     \label{Roma2}
     \and
     School of Physics and Astronomy, University of Leeds, Leeds LS2 9JT, UK
     \label{Leeds}
     \and
     Institut d'Astrophysique de Paris, CNRS (UMR7095), 98 bis boulevard Arago, 75014 Paris, France
     \label{IAP}
    \and
     University of Lyon, UCB Lyon 1, CNRS/IN2P3, IP2I, 69622 Villeurbanne, France
     \label{IP2I}
}

\abstract 
{The millimetre spectrum of Galactic regions and galaxies is rich in continuum and molecular lines. This diversity is mostly explored using either broad-band photometry or high-resolution heterodyne spectroscopy.
}
{We aim to map the millimetre continuum emission of Galactic regions with an intermediate spectral resolution between broad-band photometry and heterodyne spectroscopy, enabling us to rapidly cover large sky areas with spectroscopy.
}
{We report observations of the Orion Nebula with the CONCERTO instrument, which was installed at the APEX telescope focal plane from 2021 to 2023. 
}
{We find that the spectrum of Orion is dominated by dust emission with an emissivity index ranging between 1.3 and 2.0, along with strong CO(2-1) and H$_2$O lines, which are naturally separated from the continuum due to the CONCERTO spectral capabilities. Many regions also show strong free-free emission at lower frequencies.
}
{We demonstrate the spectral capabilities of CONCERTO at intermediate spectral resolution, with a frequency coverage from 130 to 310~GHz. A sensitivity of 200~mK is achieved in one second, for one beam and a 6\,GHz frequency width, over an 18\,arcmin diameter field of view, which is within a factor of three of the expectations. We show that we can spectrally disentangle the continuum from the CO line emission, but the line is not resolved at a resolution of $\sim 8000\ \mathrm{km s^{-1}}$. The slope of the millimetre continuum is line-free mapped for the first time in Orion. 
}
\titlerunning{Continuum, CO, and water vapour CONCERTO maps of the Orion Nebula}
\authorrunning{F.-X. Désert et al.} 
\maketitle      

\section{Introduction}\label{sec:introduction}

Dust emission pervades the (sub)millimetre sky. Composed of solid particles of submicron size, dust is a more powerful emitter than gas, although in mass it is outclassed by a factor of one hundred. In the past two decades, considerable progress has been made in our understanding of submillimeter dust emission. This progress can be attributed to contributions from \textit{Planck}, \textit{Herschel}, and ground-based observatories: Atacama Large Millimeter/submillimeter Array (ALMA), the Institut de radioastronomie millimétrique (IRAM), James Clerk Maxwell Telesecope (JCMT), Caltech Submillimeter Observatory (CSO), and the Atacama Pathfinder Experiment (APEX); see e.g. \citet{HensleyDraine2021}. Dust is the golden tracer of interstellar matter. As it reprocesses a sometimes large fraction of stellar photons, it is also widely used to measure the luminosity of objects (ranging from molecular clouds to galaxies). The dust spectral energy distribution (SED) is typically described using a modified blackbody model, with an emissivity index $\beta$ around 1.5 and temperatures ranging from 10 to 50\ kelvin (see e.g. the reviews by~\citealt{Galliano2018} and~\citealt{Hensley2023}). 
Physically motivated models, such as The Heterogeneous dust Evolution Model for Interstellar
Solids (THEMIS)~\citep{2017A&A...602A..46J} and THEMIS2~\citep{Ysard2024},  have also been proposed, taking into account the evolution of dust in the interstellar medium (ISM).
So far, dust emissivity has been probed with photometric band ratios, and little progress has been made since the Far Infrared Absolute Spectrophotometer (FIRAS) on the Cosmic Background Explorer (COBE) regarding the true underlying detailed spectrum~\citep{Boulanger1996}.
  For example, \citet{TangCMZ2021} report large variability in the dust emissivity index around the Galactic centre, and \citet{Galliano2018} describe a sub-millimetre excess emissivity in external galaxies.  Spectral millimetre mapping is thus opening a new observing space to study the SED of dust emission. Here, we present an exploratory programme designed to make the first measurement of dust millimetre emissivity with CarbON CII line in post-rEionisation and ReionisaTiOn epoch (CONCERTO) on APEX. The process of disentangling the emissivity index from temperature is relatively straightforward and is achieved by measuring the spectral index in the CONCERTO millimetre observations. Indeed, the millimetre emission occurs near the Rayleigh-Jeans limit, so the millimetre spectrum directly yields an emissivity index, with only a weak dependence on temperature (which is obtained from available \textit{Herschel} photometry).
We aim to directly test whether the emissivity law spectral index varies across different parts of star-forming regions. This will help to assess the robustness of dust-mass estimates in external galaxies and has important consequences for the dust build-up history of the Universe.

Another major motivation for measuring dust millimetre emissivity lies in the cosmological context of the search for cosmic microwave background B-modes. These polarisation modes, if detected, would establish the existence of primordial gravitational waves predicted by \textit{Planck}-favoured inflationary Big Bang models. The interstellar dust-polarised foreground is the main factor perturbing the measurements, as illustrated by the controversy over the Background Imaging of Cosmic Extragalactic Polarization (BICEP) results~\citep{BICEP2-Planck2015}. Our observations of dust emission aim to bridge the gap between small and large Galactic scales. Although CONCERTO is not a polarimeter, precise spectral measurement of dust emissivity in intensity will be highly useful for the worldwide quest for B-modes.

CONCERTO~\citep{ConcertoDesign2020, ConcertoWenkai2024} is a millimetre spectro-imager covering a field of view of 18.6~arcmin, operating over a bandpass from 130 to 310~GHz, at a spatial resolution of 27~arcseconds at 250~GHz. Its spectral resolution can reach 2.1~GHz~\citep{ConcertoWenkai2024} and it is designed for line-intensity mapping ~\citep{ConcertoBethermin2022} of the high-redshift Universe.
Here, CONCERTO is used on the bright Orion Nebula region as a benchmark, allowing existing observations to be easily compared with the new CONCERTO photometry and spectroscopy.

In Sect.~\ref{sec:obs} we present the instrument characteristics and the observations procedures.
Section~\ref{sec:data} describes the data processing, including calibration in photometric and spectroscopic modes.
Results on the millimetre spectrophotometry of Orion are presented in Sect.~\ref{sec:results} and discussed in Sect.~\ref{sec:discussion}.

\section{Observations}
\label{sec:obs}
\subsection{The CONCERTO instrument}
The CONCERTO instrument, designed in 2019 and built in 2020, was a millimetre spectral imager. It was installed at the focus of the APEX telescope in Chajnantor \citep{ConcertoDesign2020,2022EPJWC.25700010C, Monfardini2022, ConcertoWenkai2024, Fasano2022, Fasano2024} from 2021 to 2023.

\begin{figure}[!ht]
    \centering
    \includegraphics[trim={0cm 0cm 0cm 0cm},clip,width=\hsize]{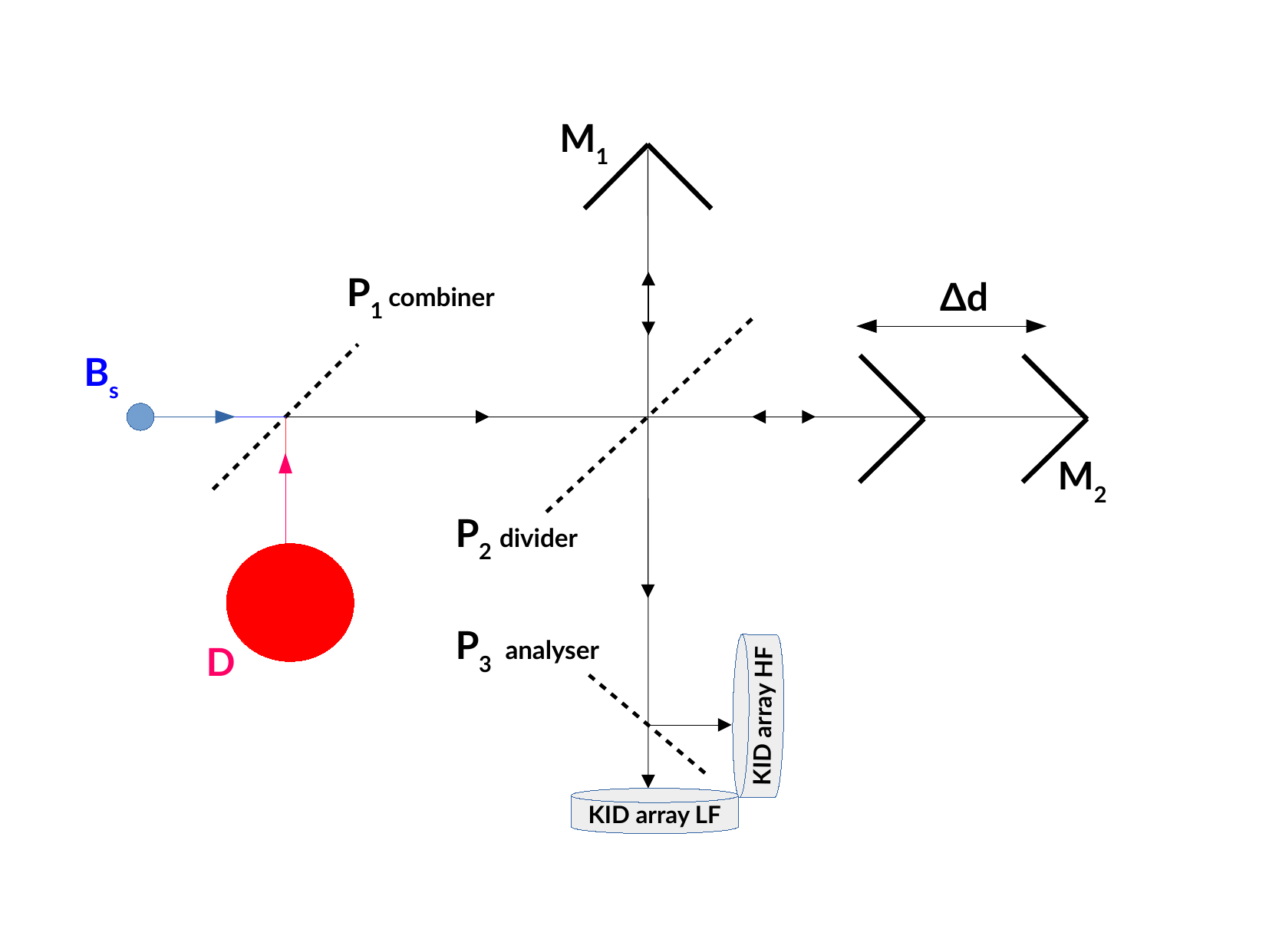}
    \caption{CONCERTO MPI optical concept. The main beams, $B_p$ and $B_s$, described in Eqs.~\ref{eq:2beamMapPhotom} and \ref{eq:2beamMapSpectro}, enter the first input port of the MPI, while an unfocussed disc ($D$) enters the second port. A first polarizing grid, $P_1$, is at the entrance of the MPI. A second grid, $P_2$,  divides the light between the two MPI arms, each ending at a rooftop mirror (two plane mirrors at 90~deg. angle), $M_1$ or $M_2$. Mirror $M_1$ is fixed, whereas $M_2$ oscillates along the horizontal axis, providing a variable OPD of $2\Delta d$. Grid $P_2$ then recombines the light and sends it into the cryostat. The light is analysed by a third polarizing grid, $P_3$, which provides the two MPI output ports feeding two KID arrays, LF and HF. Each spectral component ($\nu$) of the incoming radiation interferes constructively or destructively, depending on the moving-mirror position ($\Delta d$). The total output signal on one detector, $S(\Delta d)$, is the integral over the entire wavelength range of the input spectra, modulated by the OPD between the two arms. By performing a Fourier transform of this signal as a a function of $\Delta d$, the spectral intensity, $I_{\nu}$, of the incoming radiation can be recovered, enabling precise spectroscopic analysis.
    }
    \label{fig:MPIscheme}
\end{figure}

The instrument used about 4000~kinetic inductance detectors (KIDs) at 100~mK to map the sky with half-arcminute angular resolution.
The spectroscopic capabilities of CONCERTO were provided by a warm Martin-Puplett interferometer \citep[MPI;][]{MartinPuplett1970}, a type of Fourier transform spectrometer in which the splitting of the two beams is achieved using a polarising grid. This approach is well suited to millimetre spectroscopy. A schematic of the MPI is shown in Fig.~\ref{fig:MPIscheme}.

The figure provides details on the two input ports (discussed in Sec.~\ref{subsec:absolute}) and the two output ports, namely the low-frequency (LF) and high-frequency (HF) channels. The MPI moving mirror, $M_2$, continuously moves back and forth, producing an interferogram every $1/4\, \mathrm{s}$ for each detector. Data for all KIDs are synchronously sampled at 3815~Hz (0.26~ms).
The two output channels, each comprising about 2000~KIDs, are obtained with a cold-splitting polarising grid, $P_3$. These have different bandpasses to aid in the reconstruction of the spectra from 130 to 310~GHz, but cover the same field of view in the sky. 

\subsection{Orion observations}

This paper focusses on the analysis of the Orion Nebula (M42). This target was chosen because it is the archetype of a bright star-forming region, with the Bar serving as a template photodissociation region. It lies at a distance of $414\ \pm7\ \rm pc$ \citep{Menten2007}. The mapped region (hereafter referred to as Orion) is part of Orion~A (North).  \citet{Felli1993} showed that Orion epitomises the galactic HII region. The agreement between H$_\alpha$ and the radio emission implies that the radio emission is dominated by free-free (bremsstrahlung) emission. 

Observations with CONCERTO were conducted in 2021 and 2022, during a series of campaigns. The Orion observations described here were obtained in 2022 as the main part of the open-time ESO proposal (ID: 110.23NK, No: E-0110.C-4194A-2022) and a log summary is given in Tab.~\ref{tab:Observinglog}. The APEX scans follow the on-the-fly mapping type and are alternatively horizontal or vertical in equatorial coordinates. Eleven scans were obtained for a total integration time of 2.4 hours.
 The raster scan mode was performed with a scanning speed of 30~arcseconds per second. During mapping, the MPI rooftop mirror moved rapidly and continuously back and forth, producing a full interferogram every quarter of a second, equivalent to a 7.5~arcseconds on-sky displacement, well below the beam size. 
 Pointing and focus sequences of observations were performed regularly during Orion observations, following standard APEX procedures.

A key advantage of CONCERTO is its ability to simultaneously measure the ISM dust-emission spectrum and the molecular lines, in particular the 230\,GHz CO(2-1) rotational line. The lines are all unresolved, so the continuum spectrum is cleaned of line emission by masking the appropriate frequency bins.

\begin{table*}
\caption{Observing log for the Orion Nebula.}
\label{tab:Observinglog}
\begin{tabular}{|c|c|c|c|c|c|c|c|c|c|}
\hline
source & RA(2000) &  Dec & date  &      Obs. Time & Nb scans & Scan Duration & Size  &  pwv & elev. \\
& hr & deg. & & hr & & min & arcmin& mm & deg.\\
\hline
Orion (M42) & 05:35:17.3 & $-05$:23:28.0  & 2022-08-24  & 2.4 &    11  &     13 &         $15\times 60$  & 0.4 & 58 \\
\hline
\end{tabular}
\tablefoot{
We list the centre coordinates, the date of observation, the on-target observing time (not including the calibration time), the number of scans, the scan duration, the size of the rectangular scanning pattern  in RA and Dec., the median average precipitable water vapour (pwv), and the elevation. The maps were obtained alternately in horizontal and vertical scanning directions in equatorial coordinates.
}
\end{table*}

\section{Data analysis}
\label{sec:data}

The observed size is large compared to the field of view of CONCERTO (18.6~arcmin in diameter). The scan area must therefore be wide enough to include relatively off-source regions.
The position of the MPI moving mirror is monitored by a set of three lasers, giving the absolute linear scale of the produced interferograms. The motion is asymmetric, with a negative mechanical displacement of 30~mm and a positive one of 10~mm, relative to the zero-path-difference (ZPD) position. The optical path difference (OPD) is twice the mirror displacement. In practice, we retained only the interferogram for an absolute OPD below $OPD_{max}=2\times25\,\mathrm{mm}$. 
The maximum OPD gives a spectral resolution of $\frac{c}{OPD_{max}}= 6\ \mathrm{GHz}$. Thus, the velocity resolution is $\sim 7800\ \mathrm{km s^{-1}}\ (\frac{\nu}{230\ \mathrm{GHz}})^{-1}$. The OPD binning was chosen as $0.124\ \mathrm{mm}$ corresponding to a cutoff of the spectrum at 1200\ GHz, well above the cut-off (350~GHz) from the filters. The second input of the MPI observes an out-of-focus sky. The obtained interferogram (in addition to the continuum) is therefore the difference in brightness between the focussed main beam (about 27\ arcseconds) and an 18.6-arcminute diameter disc (the size of the field of view, centred on the same line of sight), thereby cancelling, to first order, the atmospheric emission~\citep[][]{Macias2024}.
Time-ordered information (TOI) was acquired as $I$ and $Q$ lock-in (in-phase and quadrature) detections at discrete tone frequencies, which were tuned close to KID resonances \citep{2022JInst..17P8037B}. The data were sampled simultaneously for all KIDs at a frequency of 3814.7~Hz. A round trip of the MPI rooftop mirror, including overheads, spanned a duration of 537~ms (2048 samples), which was called a block.  The average mirror scanning speed was larger than 0.2~m/s, giving $\Delta OPD \simeq 0.120\ \mathrm{mm}$  per sample. The three-point method~\citep{Fasano2021} allows recovery, from $I$ and $Q$, of the KID resonance frequency, which is assumed to be proportional to the incoming load. In practice \citep[see][]{Macias2024}, we obtained two independent datasets: a photometric timeline (photometric mode), with one average sample per block, and an interferometric timeline (spectroscopic mode), sampled at the nominal frequency of 3814.7~Hz.
The temporal efficiency (selection of samples in a block) of the spectroscopic and photometric modes was about 87\% and 95\%, respectively.
Of the 3718 KIDs read in the CONCERTO focal planes, only about 600 (LF) and 300 (HF) were used in this early spectroscopic data processing, after rejection of the others for various reasons, including severe thresholds in the quality of the TOI.

\subsection{Photometric mode}
In photometric mode, the TOI is averaged over a block and projected onto maps, one for HF and one for LF, after correction for atmospheric opacity and a common mode. The atmospheric opacity correction is computed by taking into account the CONCERTO bandpass. The bandpass of the instrument, as described by \citet{ConcertoWenkai2024}, has been measured by closing the APEX vertex, assuming blackbody radiation. 

The bandpass $H_\nu^i$ (with $i$ being LF or HF) is taken with the convention of $F \eta $ expressed by \citet{ConcertoWenkai2024} ($F_\nu$ is the relative spectral response and $\eta_\nu$ is the aperture efficiency, see their Appendix~B and Fig.~11), which is appropriate for measurements expressed in brightness temperature and for extended sources.
For each scan, the zenith opacity of the atmosphere, $\tau_\nu\mathrm{(pwv),}$ is computed with the ATM atmospheric transmission model (ATM) \citep{Pardo2001,Pardo2025},  using the precipitable water vapour (pwv) content measured simultaneously by a radiometer at APEX. The measured spectral signal for the KID $k$ of the channel $i$ is therefore

\begin{equation}\label{eq:SpectralSignal}
m_{k,\nu} \propto e^{-\mathrm{AM}\,\tau_\nu} H_{i,\nu} T_{k,\nu} \, ,
\end{equation}

\noindent where $\mathrm{AM}=1/\sin{e_l}$ is the airmass of the scan taken at elevation $e_l$, $T_{k,\nu}$ is the sky brightness temperature at the measured frequency $\nu$, assuming the Rayleigh-Jeans approximation. 

\begin{figure}[!ht]
\centering
\includegraphics[trim={0cm 0cm 0cm 0cm},clip,width=\hsize]{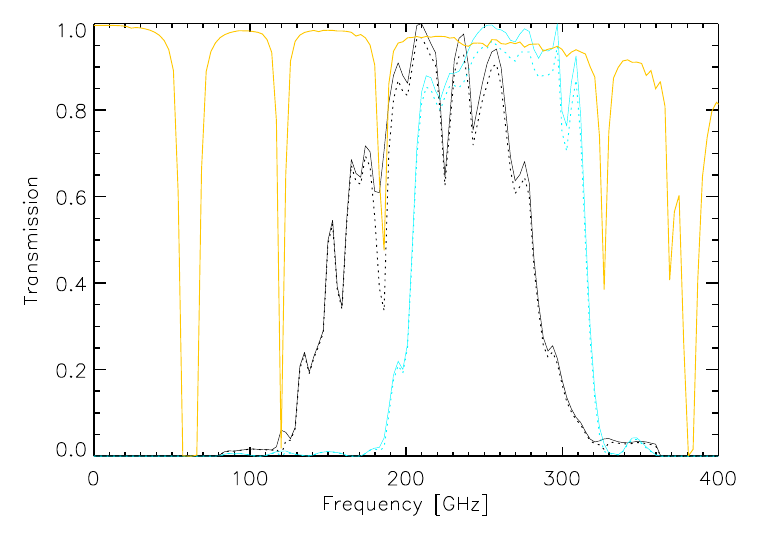}
\caption{Normalised CONCERTO bandpass for LF (black curve) and HF (green curve) channels. The atmospheric transmission is shown in orange for representative conditions during the Orion observations, with a pwv of 0.4~mm and an elevation of 60~deg. The dotted curves show the bandpasses multiplied by the atmospheric transmission. 
}
\label{fig:bandpass}
\end{figure}

Figure~\ref{fig:bandpass} shows the bandpass of the instrument, $H_{i,\nu}$, and the atmosphere during the observations. 
Assuming a typical dust spectrum of the form $T_\nu \propto \nu^{1.5}$, the atmospheric multiplicative correction factor is therefore

\begin{equation}\label{eq:AtmCorr}
c_{i} =\frac{\int  d\nu\,                         H_{i,\nu} \nu^{1.5}}
{\int  d\nu\, e^{-\mathrm{AM}\,\tau_\nu} H_{i,\nu} \nu^{1.5}} \, .
\end{equation}
The correction factor is, on average, 1.06 (for both LF and HF), with an average pwv of 0.39~mm (0.30-0.49) and AM of 1.2 over the 11 scans. It varies by at most 0.01 between scans.
The absolute calibration could have been performed with planets (see \citealt{ConcertoWenkai2024}). However, in order to use extended brightness maps, we chose to perform the calibration by comparing the CONCERTO maps with external maps (the NIKA2 maps; see below). We used masking and first-order baseline subtraction in the scanning direction to remove residual stripes in the individual scan maps. The noise map was obtained from the hit-count map and the signal-free zone statistics.

\subsection{Spectroscopic mode}
In spectroscopic mode, some additional processing is required to remove residuals in the interferograms. We followed the data model and processing presented in Section~5 of \citet{Macias2024}. We considered the LF and HF channels independently. First, we separated the raw interferograms into forwards and backwards directions. We then applied the following processing to each.
We subtracted a parasitic mirror signal induced by background modulation during the roof mirror displacement. A linear fit was subtracted from each interferogram within a block of data. We then estimated the ZPD from the mean interferogram for each detector. Next, we removed the mean interferogram per detector and per scan, which was equivalent to subtracting an overall constant per scan in the photometric mode. The bright Orion regions were masked before computing the mean interferograms. Correlated noise components were removed by applying a common mode decorrelation algorithm per scan. Finally, the cleaned interferograms for all detectors in one band were jointly projected onto the sky in OPD space using a simple co-addition method to obtain interferometric cubes. 
Combining the forwards and  backwards interferogram cubes and taking the fast Fourier transform (FFT), without apodisation\footnote{Apodisation  is commonly performed by tapering the interferogram at its high OPD end to achieve a smoother spectral response, at the cost of spectral resolution.}, we obtained spectral cubes for the LF and HF bands. We considered the same number of OPD bins for all scans so that the normalisation of the FFT could be accounted for in the calibration. 

As in photometric mode, we used masking and first-order baseline subtraction in the scanning direction to remove residual stripes for each individual spectral slice of the scan maps. The statistical noise (used only for weighting\footnote{The cubes coming from different scans were co-added with a weighting in $1/\sigma^2$, where $\sigma$ is the statistical noise.}) was assumed to be spectrally flat (see examples of noise spectra in \citealt{ConcertoWenkai2024}) and was measured in two clean spectral windows between 450 and 500~GHz, and 650 and 700~GHz for both LF and HF, where the bandpass is zero. 
The line profile is addressed in Eq.~\ref{eq:EmLine}, yielding a fixed spectral resolution of 6~GHz, sampled at 3~GHz.

\subsection{Absolute calibration}\label{subsec:absolute}
We calibrated the CONCERTO LF and HF channels using an external dataset provided by the New IRAM Kids Arrays (NIKA2)~\citep{Perotto2020} and acquired in the same region. The NIKA2 instrument is a KID-based camera with many instrumental features in common with CONCERTO, including the KID read-out electronics. This camera has an instantaneous field of view of 6.5', covered at two simultaneous frequencies of 260 and 150~GHz, with an angular resolution of 12.5 and 18.5".

An additional difficulty for calibration is the presence of two beams on the sky, which contribute equally to the CONCERTO map when the sky-sky interferogram configuration is used (the case throughout this paper). Two input ports are involved: one with the nominal beam and the second being an 18.6~arcmin disc-like unfocussed beam (this second input is described in Sect.~\ref{sec:obs}). In photometric mode, once calibrated, CONCERTO measures a map $T^{mp}$ of the sky temperature\footnote{The Rayleigh-Jeans law numerically links the brightness temperature to the intensity with $(B_\nu/\mathrm{1\ MJy/sr}) = 0.0307\ (\nu/\mathrm{1\ GHz})^2\ (T/\mathrm{1\ K})$} as

\begin{equation}\label{eq:2beamMapPhotom}
T^{mp} = T*B_p + T*D\, ,
\end{equation}
\noindent where the convolution ($*$) of the sky temperature $(T)$ is achieved with the sum of $B_p$, the photometric beam response (as described by \citealt{ConcertoWenkai2024}), with a full width at half maximum (FWHM) of 32.2 and 28.6" for LF and HF (hereafter approximated as 30") of integral unity, and $D$, the CONCERTO field of view, a centred and uniformly illuminated disc of integral unity. The ratio of the average photometric beam ($2\times 10^{-8}\,\mathrm{sr}$) to the disc ($2.3\times 10^{-5}\,\mathrm{sr}$) solid angles is approximately $10^{-3}$. Consequently, the flux of a compact source is negligibly affected by the disc. On the contrary, a nearly flat brightness will translate into a factor two miscalibration, because either the planet or NIKA2 calibration is for a single nominal beam. 

In spectroscopic mode, the two ports are differenced instead of summed. This feature provides an efficient sky-noise cancellation. The measured map for each frequency, $\nu$, is

\begin{equation}
\label{eq:2beamMapSpectro}
T^{ms}_\nu = T_\nu*B_s^\nu - T_\nu*D\, ,
\end{equation}where $T_\nu$ is the sky temperature at frequency $\nu$. The nominal beam in spectroscopy $B_s^\nu$ is measured (on Mars) to be at the diffraction limit, with the FWHM being 27~arcsec$\times (250\,\mathrm{GHz}/\nu)$, an expression valid for diffraction on an entrance pupil aperture of 11~m diameter (compared with the full APEX 12~m diameter). We discuss the use of Eqs.~\ref{eq:2beamMapPhotom} and \ref{eq:2beamMapSpectro} in the next subsection. 

\subsubsection{NIKA2 Orion map}

The NIKA2 data on Orion were taken as part of the polarisation large programme `Probing the B-field in star-forming Filaments Using NIKA2-Pol' (B-FUN; programme  015-17, P.I. Ph. André) at the IRAM 30~m telescope (near Pico Veleta, Granada, Spain). Twenty-one scans were reduced from observations from 6, 7, 11, and 14 February 2022. The 11.3~minute scans were in equatorial coordinates, with a size of 11x11 arcmin, an orientation alternating between 0, $-45$, and $+45$~deg, a speed of 40~arcsec/s, and two different pointings. Data were taken at a 48~Hz sampling adapted for polarisation measurements with a continuously rotating half-wave plate with a spin rate of 3~Hz. Here, we used only the NIKA2 Stokes $I$ co-added map at 260~GHz, as the central frequencies of LF and HF are close to this value (225 and 262~GHz, ~\citealt{ConcertoWenkai2024}).  NIKA2 data were processed with very mild filtering options, so diffuse emission was preserved. A mask was used around the central Orion emission and atmospheric de-correlation was performed with a template composed of KIDs outside the mask. The map was originally calibrated in Jy/beam using the Uranus flux model at 260~GHz. We then multiplied by a colour correction (1.15), computed with the NIKA2 bandpass, to correct to a fiducial $I_\nu \propto \nu^{3.5}$ spectrum (in brightness). We then divided by the NIKA2 beam solid angle of 211~arcsec$^2$ to convert the NIKA2 260~GHz map into a dust-like brightness temperature at a reference frequency of 250~GHz (the overall factor from Jy/beam to K being $0.15$, including an estimated filtering correction factor of $1.4\pm0.2$ appropriate for spectroscopy). This NIKA2 map is an intermediate product of the B-FUN programme (André et al., in prep.).
In order to compare the NIKA2 map (re-scaled to 250~GHz) to CONCERTO maps (LF and HF, also re-scaled to 250~GHz), we convolved them all to a common final angular resolution of 40~arcseconds and a common astrometry. This value was chosen so that the noise could be neglected in the following correlation analysis. We also co-added LF and HF maps. 

\subsubsection{Photometric calibration}
\begin{figure*}[!ht]
\centering
\includegraphics[trim={0cm 0cm 0cm 0cm},clip,width=0.3\hsize]{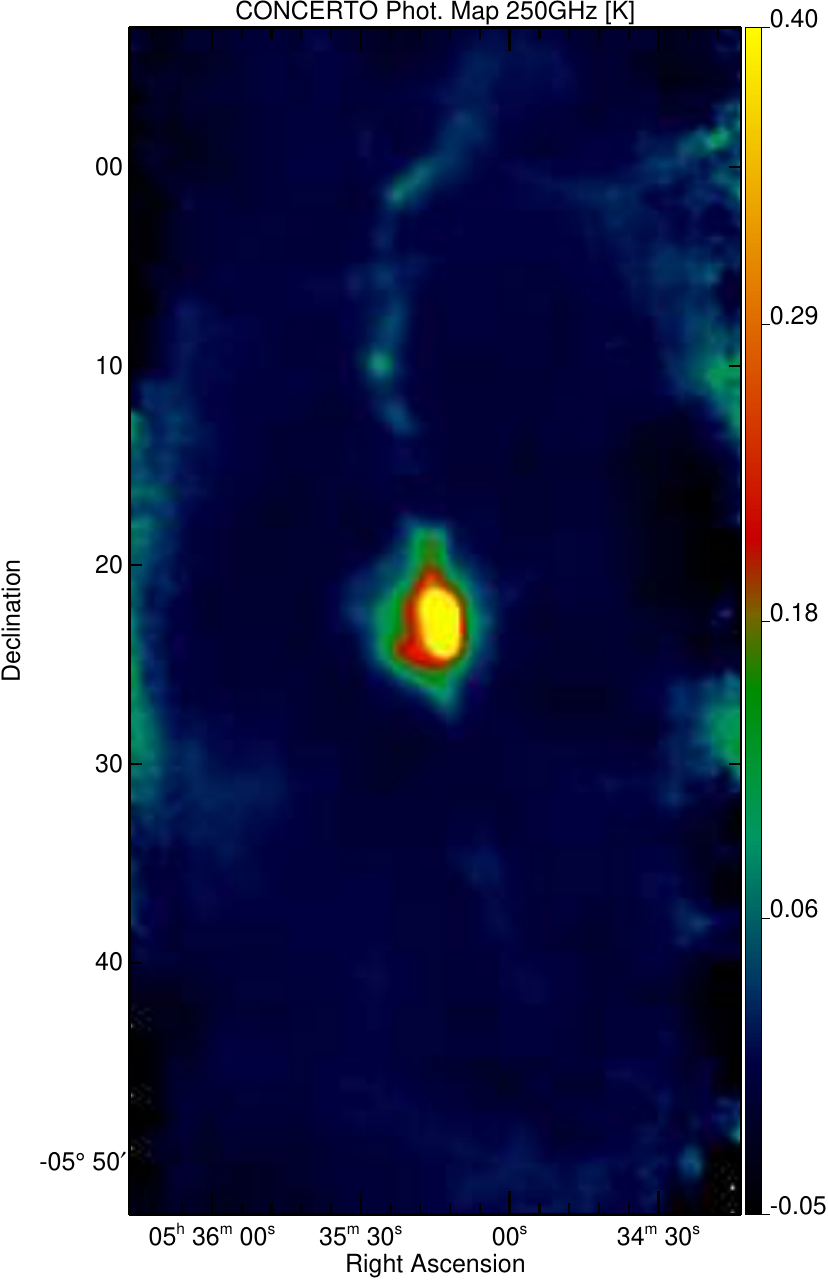}
\includegraphics[trim={0cm 0cm 0cm 0cm},clip,width=0.3\hsize]{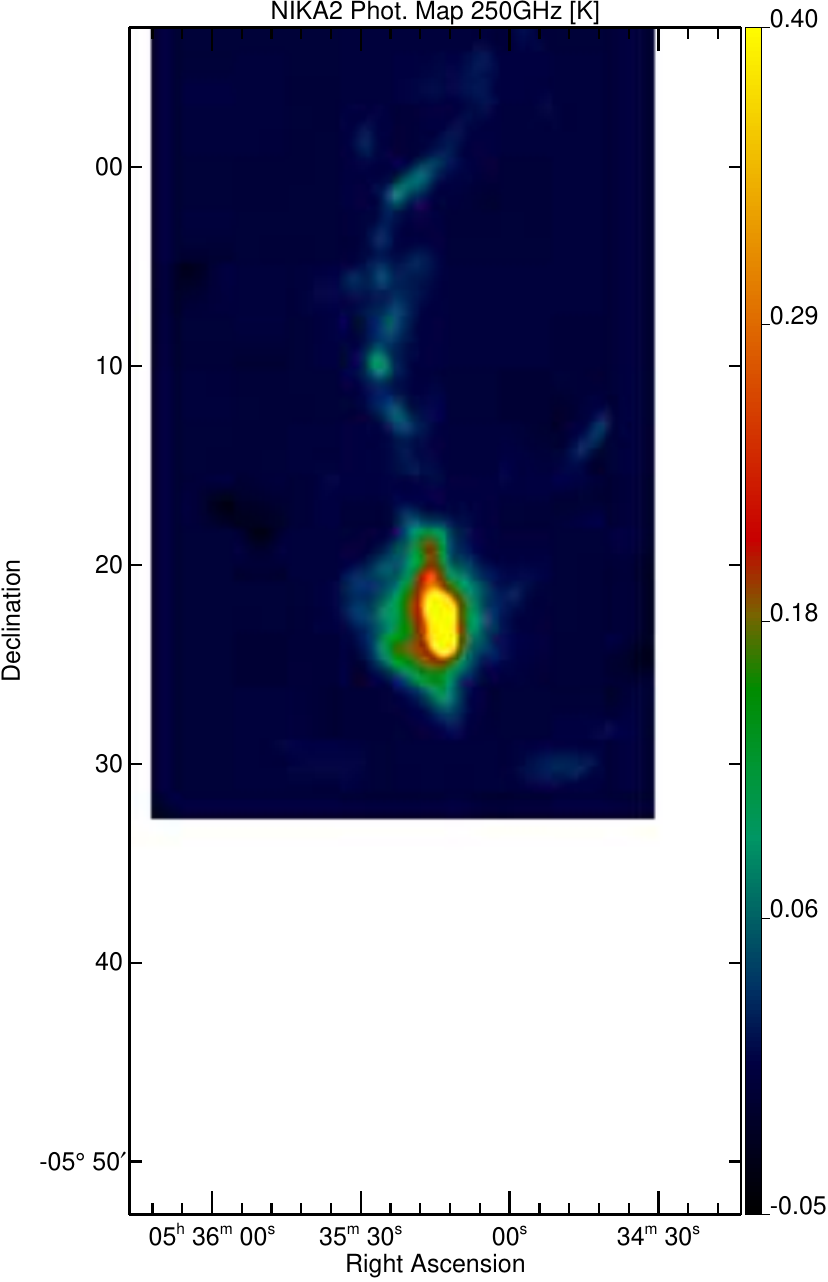}
\caption{Photometric map of Orion with CONCERTO (\textit{left}) and NIKA2 (\textit{right}) in brightness temperature at 250~GHz, convolved to a common 40~arcsec resolution. The colour bar is in kelvins. The map is saturated at 0.4~K to enhance faint features.
}
\label{fig:Map_PhotCTO_NIKA2}
\end{figure*}

\begin{figure}[!ht]
\centering
\includegraphics[trim={0cm 0cm 0cm 0cm},clip,width=\hsize]{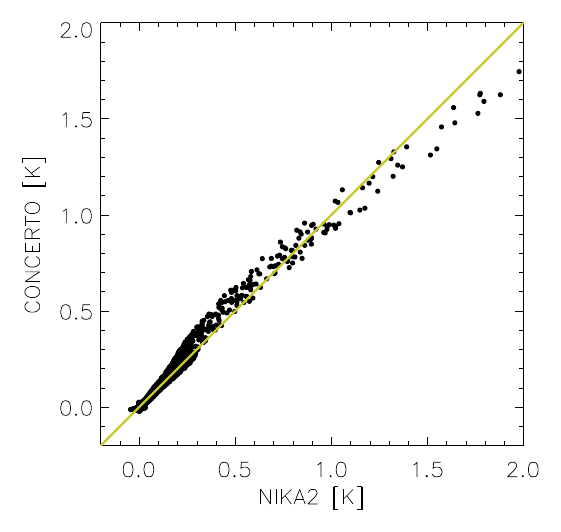}
\caption{Pixel-by-pixel correlation of the photometric brightness temperature of Orion as measured by CONCERTO with respect to NIKA2, both at 250\,GHz, as shown in Fig.~\ref{fig:Map_PhotCTO_NIKA2}. The pixel size is 10~arcseconds and the angular resolution is 40~arcseconds. The CONCERTO data are calibrated using Eq.~\ref{eq:2beamMapPhotom}. The yellow line has a slope of unity.
}
\label{fig:Corr_PhotCTO_NIKA2}
\end{figure}

We first inter-calibrated all KIDs of the same channel. For the CONCERTO photometric calibration, we then computed a single factor using Eq.~\ref{eq:2beamMapPhotom}, with $T$ taken from the NIKA2 data. We applied the same filtering to the NIKA2 data as that induced by the two beams and by the CONCERTO de-striping.
The impact of the disc term in Eq.~\ref{eq:2beamMapPhotom} is at the 4~\% level only. 
The two processed and calibrated maps (CONCERTO and NIKA2, scaled and brought to the same angular resolution) are shown in Fig.~\ref{fig:Map_PhotCTO_NIKA2}. Fig.~\ref{fig:Corr_PhotCTO_NIKA2} shows the pixel-to-pixel comparison of CONCERTO maps to the NIKA2 map. There is a satisfactory agreement in the correlation, although filtering residual effects might induce small deviations from linearity that can be seen at intermediate brightness. The CONCERTO photometric maps can be deemed absolutely calibrated by using NIKA2, with an absolute error of roughly 15 percent (taking into account the colour correction and beam approximations). We find that the typical response of KIDs is $1.5\,\rm mK/Hz$ (for LF and HF at 250~GHz).

We further verified that the CONCERTO calibration with NIKA2, as an extended calibration, is consistent with the point-source CONCERTO calibration on Uranus and quasars performed by~\citet{ConcertoWenkai2024}, provided that the conversion uses a solid angle close to the main beam of the CONCERTO instrument. 
We also checked the 260~GHz NIKA2 calibration. and found it consistent with \textit{Planck} satellite maps, after applying proper frequency conversion, colour correction, and brightness scaling by $\nu^{3.5}$, using the \textit{Planck} nominal beam at 217~GHz. For Orion, we find a ratio of 0.95 between NIKA2 and \textit{Planck}, which is consistent with NIKA2's absolute photometric precision. 
Photometric noise was measured from the scatter in pixels away from the main Orion emission. We find that if the noise scales as the inverse of the square root of time, the constant is $31\,\rm mK\,s^{\frac{1}{2}}$ per beam, a value 15 times higher than reported by~\citet{ConcertoWenkai2024}.
We suspect that this was due to contamination by residual emission rather than instrumental noise in the so-called empty measurement area. 

\subsubsection{Spectral calibration}
The spectral LF and HF projected cubes in raw units were calibrated using the corresponding NIKA2 photometric maps (matching Eq.~\ref{eq:2beamMapSpectro} and the de-striping). For that purpose, we used the bandpasses $H_\nu^i$ (with $i$ being LF or HF), which can be regarded as a spectral flat field for the cubes. From Eq.~\ref{eq:SpectralSignal}, we rewrote the measured spectral signal as
\begin{equation}\label{eq:SpectralSignal2}
m_{i,\nu} = k_i\, e^{-\mathrm{AM}\,\tau_\nu} H_{i,\nu} T^{ms}_\nu \, ,
\end{equation}
where $k_i$ is the calibration factor to be determined (independent of frequency), with one factor per channel (LF or HF). In this spectroscopic calibration method, we forced the calibration factor so that the cube, integrated over frequencies, matched the NIKA2 map filtered like CONCERTO.
We then deduced the absolutely calibrated cube with
\begin{equation}\label{eq:SpectralSignal3}
T^{ms}_\nu = m_{i,\nu} / (k_i\, e^{-\mathrm{AM}\,\tau_\nu} H_{i,\nu})  \, ,
\end{equation}
applying the same factor to the noise cube. Naturally, the noise becomes infinite where the bandpass or the atmosphere has a zero-transmission (in practice, below 130~GHz and above 310~GHz).

\section{Results on the Orion Nebula}
\label{sec:results}

\begin{figure}[!ht]
\centering
\includegraphics[trim={0cm 0cm 0cm 0cm},clip,width=\hsize]{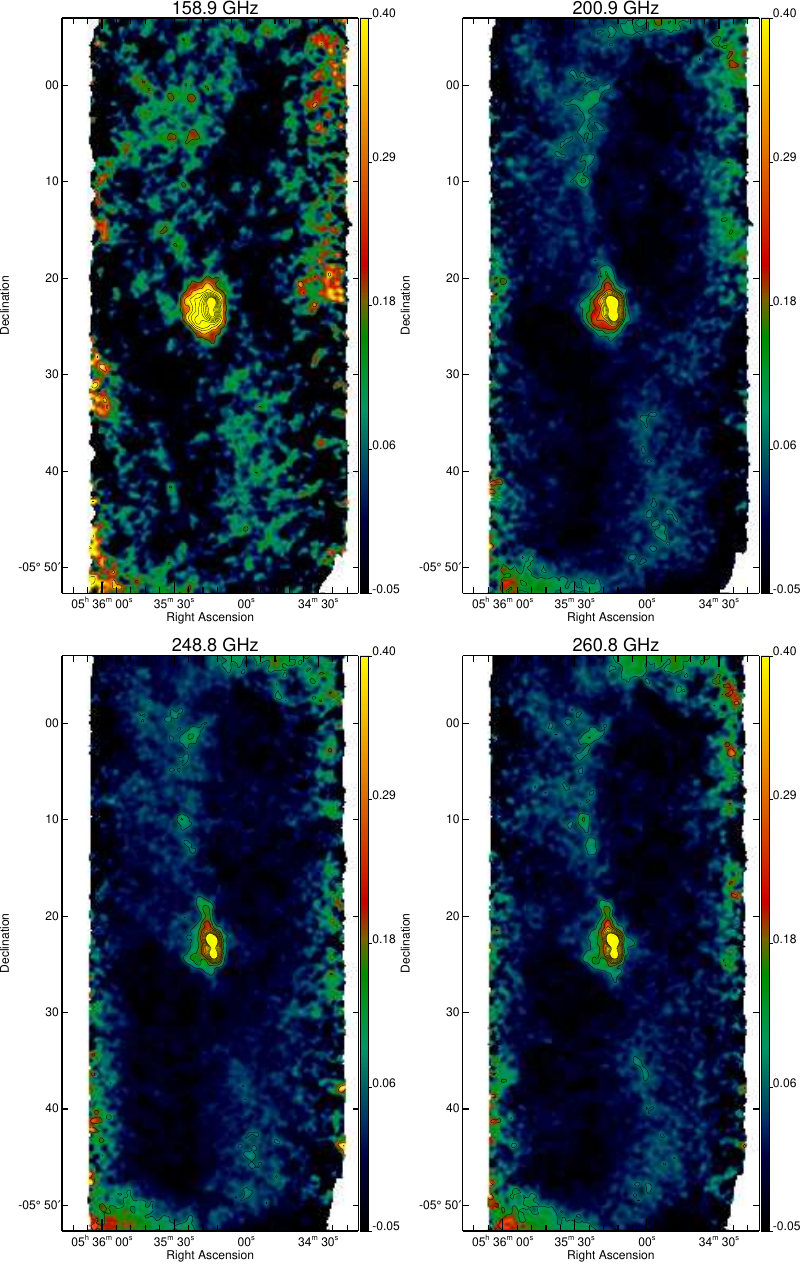}
\caption{Slices of Orion temperature brightness at specific frequencies, in 3~GHz bins, across the CONCERTO bandpass. Each map is convolved with a 30~arcsec Gaussian. Contours are in increments of $3\,\sigma$.
}
\label{fig:OrionSlice}
\end{figure}

Using the calibration and de-striping scheme described above, we obtain a clean, calibrated spectral cube. In Fig.~\ref{fig:OrionSlice}, we show a sample of slices across that cube at different frequencies. At low frequencies, the radio emission (free-free and synchrotron) is dominant, while dust emission is increasingly important at higher frequencies.

\begin{figure*}[!ht]
\centering
\includegraphics[trim={0cm 0cm 0cm 0cm},clip,width=0.24\hsize]{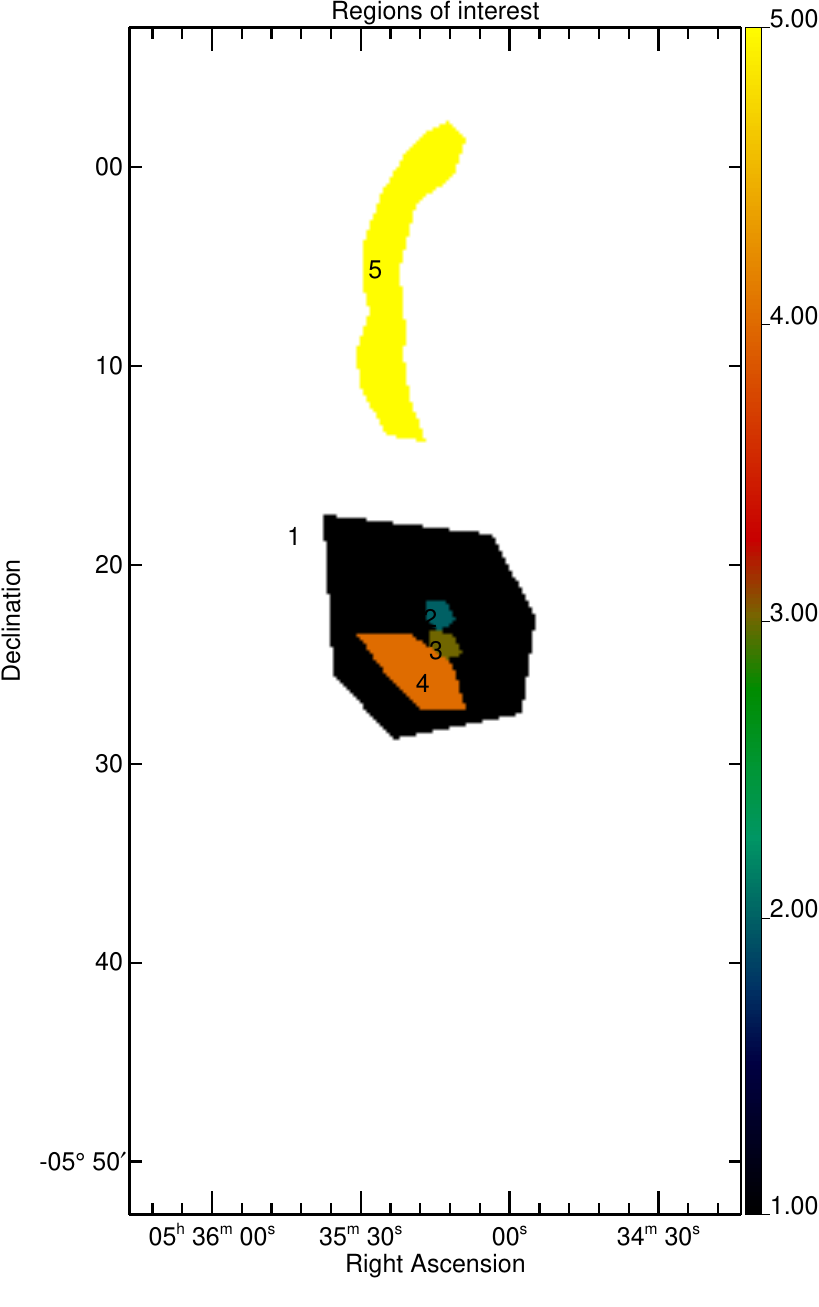}
\includegraphics[trim={0cm 0cm 0cm 0cm},clip,width=0.7
\hsize]{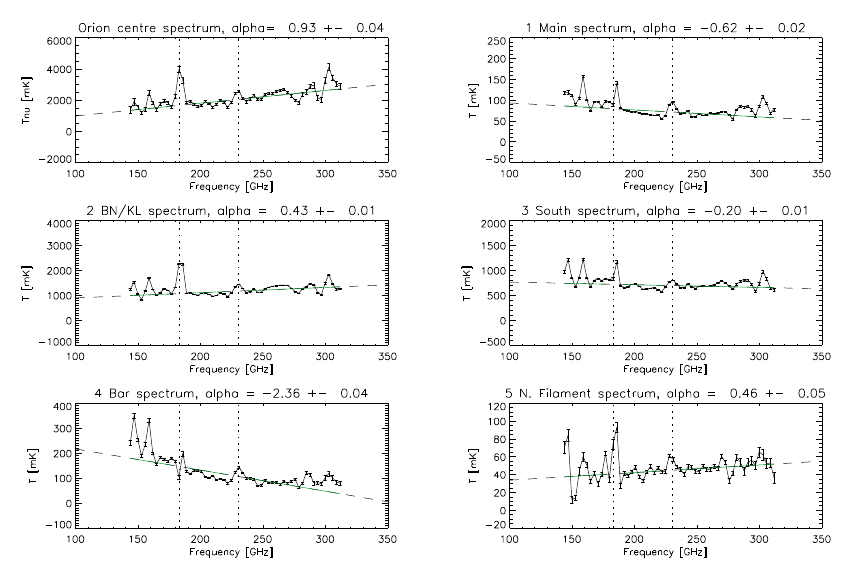}
\caption{Spectrum of selected regions in Orion. The left figure shows the selected regions in equatorial coordinates with their labels. The average spectra of these regions are presented in the right panels, the first of which corresponds to the spectrum of the 10~arcsec  pixel of the Orion peak at 5:35:14 -5:22:30. The green curve is a linear fit to the continuum from 144 to 312~GHz, excluding the CO and H$_2$O lines at 230.5 and 183.3~GHz. 
The resulting power-law index $\alpha$ at 250~GHz (as defined in Eq.~\ref{eq:alpha}) is given in the panel title. The values differ from those in Tab.~\ref{tab:roi} because of the different frequency range. The CO(2-1) line is clearly detected at 230.5~GHz, and the water-vapour line is also apparent at 183.3~GHz, as indicated by the two vertical dotted lines). 
}
\label{fig:OrionSpectrum}
\end{figure*}

We measured the average spectrum in designated sky regions of interest. Fig.~\ref{fig:OrionSpectrum} shows the variety of spectra obtained. 

In addition to continuum emission, the CO(2-1) line is present in all spectra. We decomposed the emission spectrum into a continuum component and the CO line as follows:
\begin{equation}\label{eq:EmDecomp}
T_\nu=T_\nu^c + T_\nu^{CO},
\end{equation}
where we approximated the continuum as a linear spectrum around the reference frequency $\nu_0=250\ \mathrm{GHz}$:
\begin{equation}\label{eq:EmCont}
T_\nu^c = T_0+ \Delta T \ \frac{\nu-\nu_0}{\nu_0},
\end{equation}
so that an underlying power-law spectrum has an index which is obtained by a Taylor expansion as
\begin{equation}\label{eq:alpha}
    \alpha=\Delta T/T_0\,.
\end{equation}
For a dust spectrum in the Rayleigh-Jeans regime (i.e. at high dust temperatures), the index $\alpha$ is equal to the dust emissivity index $\beta$. The parameters $T_0$ and $\Delta T$ can be solved with a basic linear fitting, including the weights. The weights are derived from the inverse variance of the signal (outside Orion) and the integration time per pixel. 

The CO line is unresolved and modelled as
\begin{equation}\label{eq:EmLine}
T_\nu^{\rm CO} = \frac{\nu_{\rm CO}}{c}\ I_{\rm CO}\ g_{ILS}(\nu)
,\end{equation}
where $g_{ILS}$ is the profile (with an integral of unity) corresponding to the integrated line shape (\citealt{Naylor2014}), and $I_{\rm CO}$ is the velocity-integrated CO line intensity (in $\mathrm{K\,km\,s^{-1}} $). Once we have obtained the continuum (Eq.~\ref{eq:EmCont}), we can subtract it from Eq.~\ref{eq:EmDecomp}, and fit for $I_{\rm CO}$ using a least-square method applied to Eq.~\ref{eq:EmLine}.
We further define the dimensionless equivalent width (EW) of the line as

\begin{equation}\label{eq:EW}
EW = \int \frac{d\nu}{\nu_{\rm CO}}\ \frac{T_\nu - T_\nu^c}{T_\nu^c}= \frac{I_{\rm CO}}{c\ T_{\rm CO}^c}.
\end{equation}

\noindent The EW quantifies the extent to which a line can contaminate a continuum measurement in photometric observations. Typically, a flat broad-band $\Delta\nu$ measurement is perturbed by the line by a fraction $ EW\ \frac{\nu_{\rm CO}}{\Delta\nu}$.

\subsection{Continuum study}
As expected, we observe regions with a clear dust-like emission spectrum increasing with frequency and others with a radio-like emission decreasing with frequency. This diversity suggests that our results are not affected by major systematic residuals. 
Fig.~\ref{fig:OrionSpectrum} (right) displays the millimetre spectrum of selected regions defined in Fig.~\ref{fig:OrionSpectrum} (left). 
The frequency limits of these spectra have been set at 144 and 312~GHz where, outside that window, the noise per spectral bin is more than four times greater than the minimum noise.
The spectra are dominated by the continuum emission. A prominent emission line is visible at 230.5~GHz, which is identified as the rotational transition of CO(2-1) (at 230.542~GHz).
An H$_2$O line is also visible at 183.3~GHz (see below).

\begin{figure*}[!ht]
\centering
\includegraphics[trim={0cm 0cm 0cm 0cm},clip,width=\hsize]{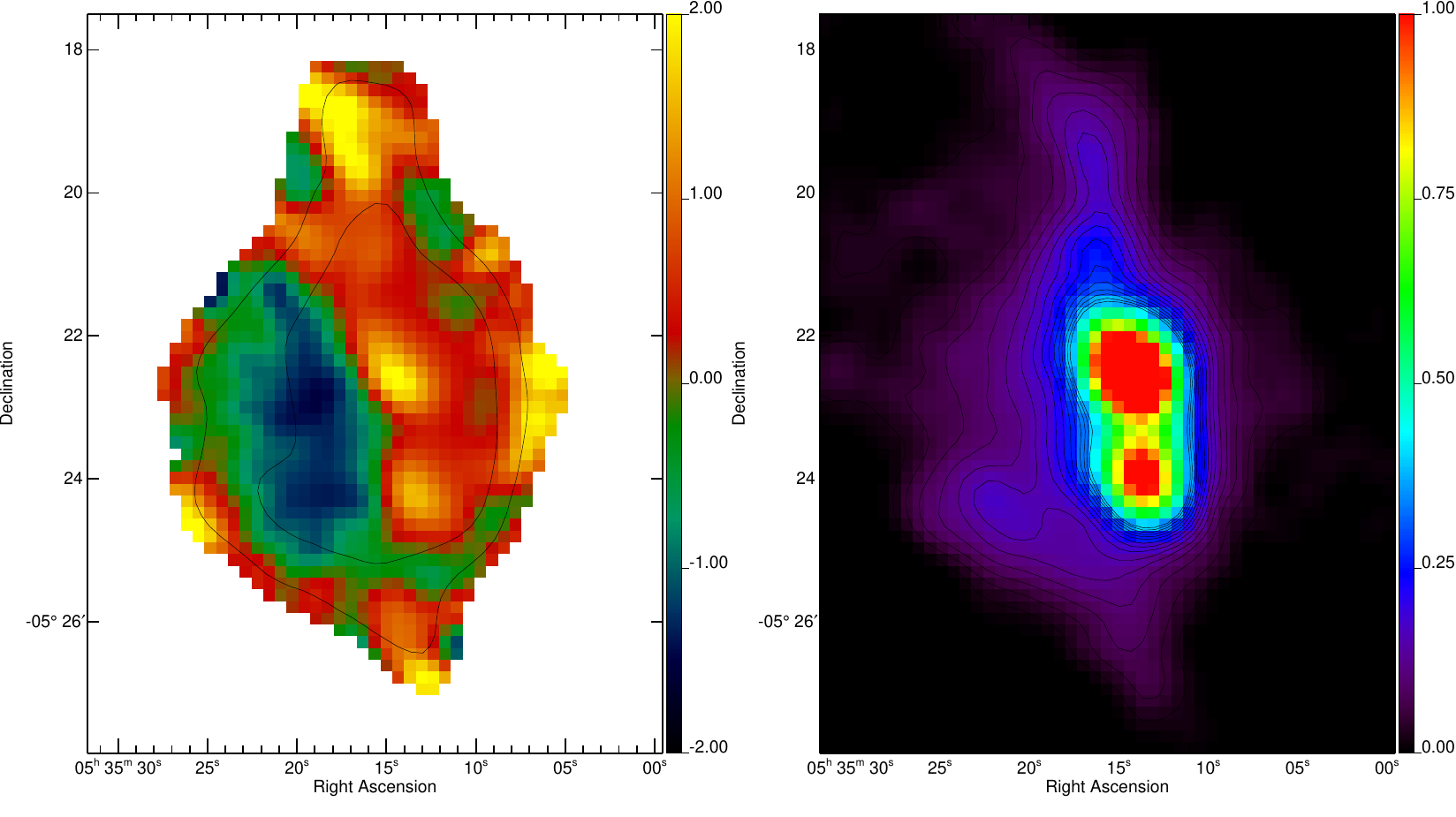}
\caption{
Left: Map of the power-law index $\alpha_{250}$, centred at 250~GHz, of the continuum  brightness temperature in Orion (see Eq.~\ref{eq:EmCont}), after convolution to a resolution of 40~arcseconds, focussing on the Orion Centre. The power-law index is uncontaminated by the CO line. 
Contours indicate uncertainties of 0.3 (inner) and 0.6 (outer) for  $\alpha_{250}$ within one beam. The map is clipped at an uncertainty of 1. 
Right: Map of the 250~GHz continuum in kelvins at a 40~arcsec resolution. The S/N is labelled on the contours in steps of 5, with a typical noise of 6~mK.
}
\label{fig:OrionAlpha}
\end{figure*}

We note that the continuum is clearly a mixture of the expected components: dust and free-free emission. The maximum index, $\alpha$, is barely above unity, so dust emission is clearly not the only emission process in Orion. The overall average spectrum of Orion has an index of $\alpha_{200} \sim -2$, which indicates that it is dominated by free-free emission. At higher frequencies, dust emission becomes dominant (see the $\alpha_{260}$ column in Tab.~\ref{tab:roi}).
The main continuum result of this analysis is shown in Fig.~\ref{fig:OrionAlpha}, which presents the spectra shown earlier in a different format. The Orion Centre is dominated by a positive index, $\alpha_{250}$ (defined between 210 and 290~GHz), of around 1, while the bar shows a negative (free-free) index of around $-2$. 

\begin{figure}[!ht]
\centering
\includegraphics[trim={0cm 0cm 0cm 0cm},clip,width=\hsize]{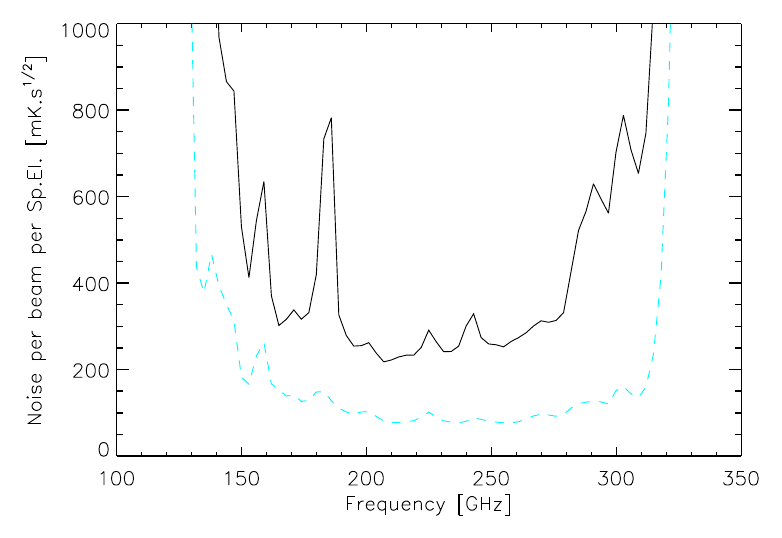}
\caption{Spectral sensitivity of CONCERTO.  For a one second integration and a reference beam of 30~arcsec FWHM, the noise level is shown as a function of frequency (solid black curve) per spectral element of 3~GHz, for an average pwv of 0.4\,mm.
The dashed cyan curve represents the instrumental noise, estimated with the measured noise outside the CONCERTO bandpass, extrapolated to CONCERTO frequencies assuming white noise, and converted to physical units using the spectroscopic response without atmospheric correction. 
}
\label{fig:OrionSpSens}
\end{figure}

\subsection{CONCERTO spectral sensitivity}
Figure~\ref{fig:OrionSpSens} shows the spectral sensitivity obtained for Orion. The noise level is measured as the rms fluctuations of each individual frequency slice outside the main Orion signal, which is then scaled to one second of integration time. This may be an overestimate of the noise, as some residual sky signal could still contribute to the rms. The sensitivity curve corresponds to the average pwv and elevation of the present observations. The atmospheric opacity around the water vapour line at 183\,GHz has a strong influence on the sensitivity around this frequency (see Fig.~\ref{fig:bandpass}). We can also extrapolate the sensitivity from the instrumental noise measured in the Hz internal KID response units at high frequencies (450-500~GHz and 650-700~GHz), where no signal is expected due to the CONCERTO bandpass filtering. Assuming a flat instrumental noise, we can then divide it by the absolute response (which contains the bandpass without atmospheric correction), and obtain the dashed cyan curve, which is a factor of 2.5-3 times lower than the black curve. We note that the bandpass features appear inverted, particularly at 160, 185, 225, and 245~GHz. The 160~GHz feature is possibly an effect of the cut-on of HF.

The median measured spectral noise (from 200 to 280~GHz) is  $268\, \rm mK\,s^{\frac{1}{2}}$ per beam for a 3~GHz spectral bin. This corresponds to an effective time, i.e. during the useful part of the interferogram. 
On average, a 10"-square pixel is observed for 33\,s by KIDs, so the noise in the fully covered area is around $47\,\rm mK$ per spectral bin of 3~GHz, per beam. The relatively low integration time per beam is due to restrictive KID selection, which will be improved in future studies.\\ 

From the dashed curve in Fig.~\ref{fig:OrionSpSens}, we estimate a median instrumental noise measurement (from 200 to 280~GHz) of about $88\, \rm mK\,s^{\frac{1}{2}}$ per beam for a 3~GHz spectral bin. This matches expectations based on the CONCERTO photometric sensitivity of $1-2\, \rm mK\,s^{\frac{1}{2}}$ and accounts for spectroscopic efficiency and the Martin-Puplett factor (close to the number of spectral elements, here 57).

\subsection{Study of the CO line }

We detected intense CO(2-1) line radiation from the direction of the Orion Nebula at a frequency of 230.538~GHz~\citep{CDMS_Endres2016}. Fifty-five years ago, carbon monoxide was first detected at the 115,267.2~MHz (1-0) transition by~\citet{Wilson1970} in this same nebula.

\begin{figure*}[!ht]
\centering
\includegraphics[trim={0cm 0cm 0cm 0cm},clip,width=0.6\hsize]{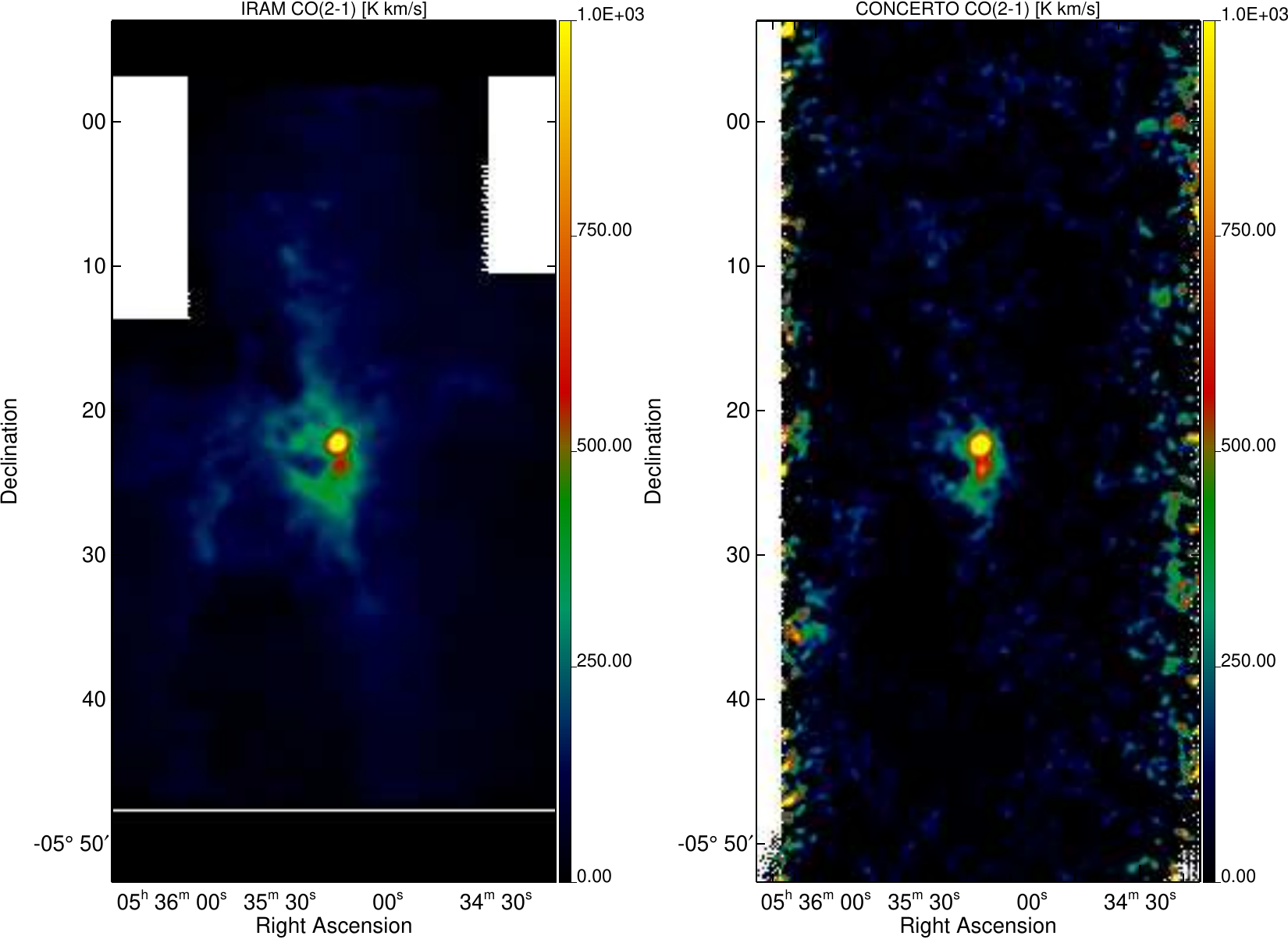}
\caption{Left: Integrated CO(2-1) line emission map from \citet{Berne2014}, obtained with the IRAM 30m telescope, in $\mathrm{K\,km\,s^{-1}}$,  re-projected into CONCERTO coordinates after convolution to an effective FWHM of 40~arcseconds. Right: Integrated CO(2-1) line map measured by CONCERTO. The integrated line intensity (in $\mathrm{K\,km\,s^{-1}}$) is derived using Eq.~\ref{eq:EmLine}, after subtracting the continuum. The map is convolved with a 30~arcsec Gaussian kernel to reach an effective FWHM of 42~arcseconds.}
\label{fig:OrionCO_IRAMConcerto}
\end{figure*}

\begin{figure}[!ht]
\centering
\includegraphics[trim={0cm 0cm 0cm 0cm},clip,width=\hsize]{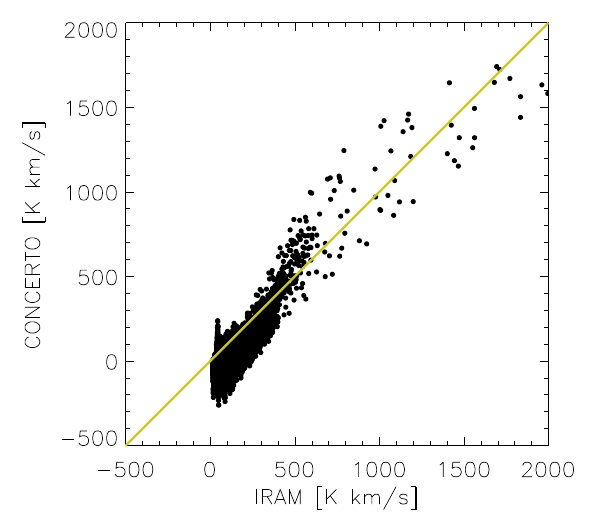}
\caption{Pixel-by-pixel correlation of Orion CO(2-1) integrated line brightness measured by CONCERTO with respect to the IRAM 30~m heterodyne map from \citet{Berne2014}. The pixel size is 10~arcseconds. The yellow line has a slope of unity. The typical noise for the CONCERTO map is $96\,\mathrm{K\,km\,s^{-1}}$
}
\label{fig:Corr_CTO_Berne}
\end{figure}

Fig.~\ref{fig:OrionCO_IRAMConcerto} shows the line-integrated intensity of CO(2-1) in Orion at the same frequency, with two completely independent datasets: the IRAM 30~m map (left) by \citet{Berne2014}, and the CONCERTO/APEX map (right) computed here.
Fig.~\ref{fig:Corr_CTO_Berne} shows the correlation between the two datasets, with the IRAM 30\,m map on the horizontal axis and CONCERTO/APEX on the vertical axis. The quantitative agreement is remarkable. We highlight that the observation time by \citet{Berne2014} was 25~hours. Due to the CONCERTO spectral resolution used, we have not attempted to separate the main $^{12}$CO line from the $^{13}$CO line at 220.399~GHz~\citep{CDMS_Endres2016}. Using the IRAM 30\,m observations in both lines~\citep{Berne2014}, we see that this isotopic line could perturb the main line measurement at the ten-percent level. 

\begin{figure}[!ht]
\centering
\includegraphics[trim={0cm 0cm 0cm 0cm},clip,width=\hsize]{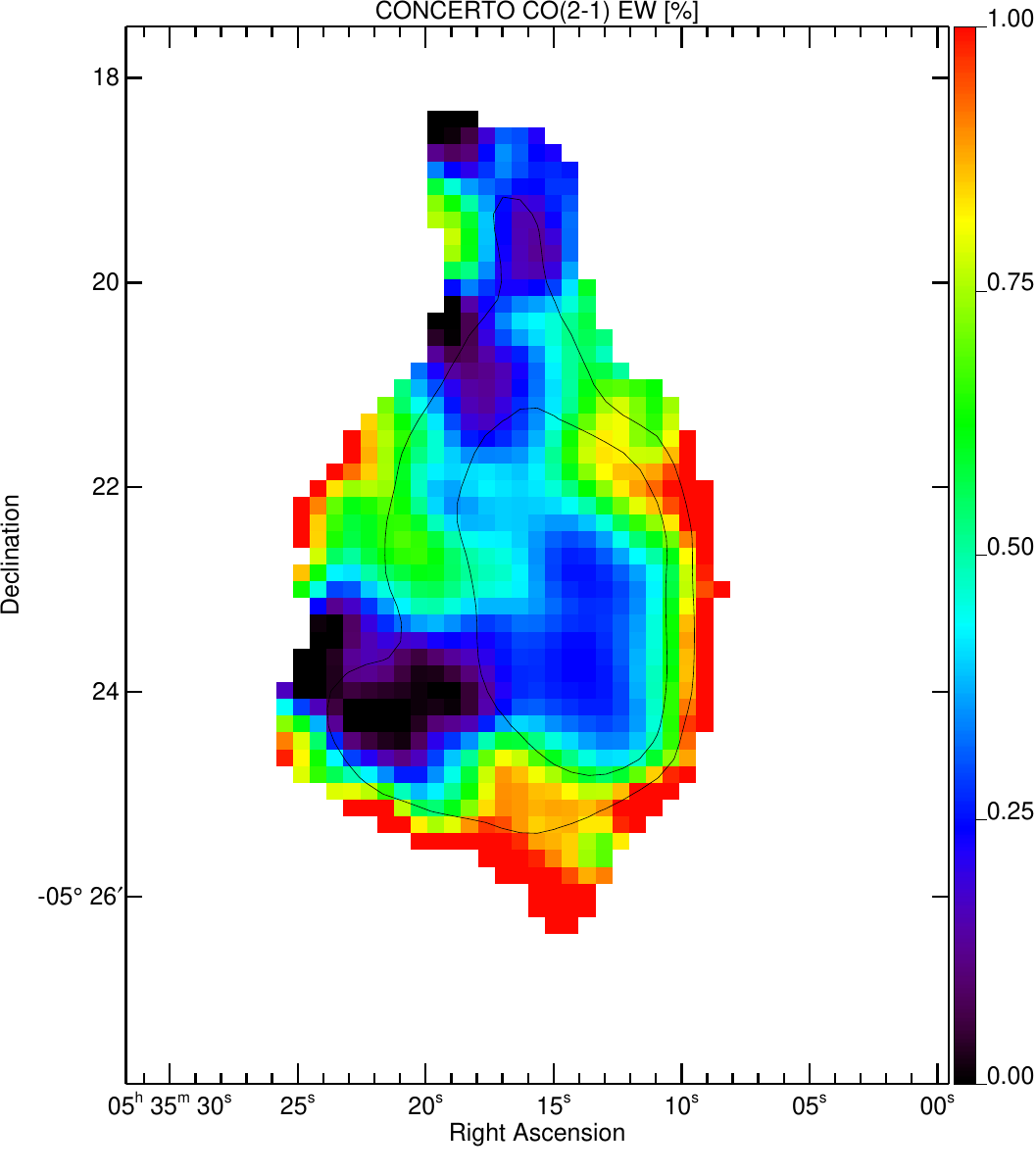}
\caption{EW map of the Orion CO(2-1) line, in percent, as defined by Eq.~\ref{eq:EW}. This is a dimensional measure of the CO contribution to the spectrum. Contours indicate the noise at $0.1$ and $0.2\%$, increasing outwards. 
}
\label{fig:OrionEW}
\end{figure}

Figure \ref{fig:OrionEW} shows the Orion map of the CO EW,  defined in Eq.~\ref{eq:EW}. The centre of Orion has an average $EW$ of $3\times10^{-3}$, indicating that in a photometric measurement with $R=\nu/\Delta\nu=3$, the CO(2-1) would contribute to approximately one percent of the continuum ($EW/R$).
The $EW$ is even smaller on the Orion Bar, at around $1\times10^{-3}$. 

\subsection{H$_2$O line study}

\begin{figure}[!ht]
\centering
\includegraphics[trim={0cm 0cm 0cm 0cm},clip,width=\hsize]{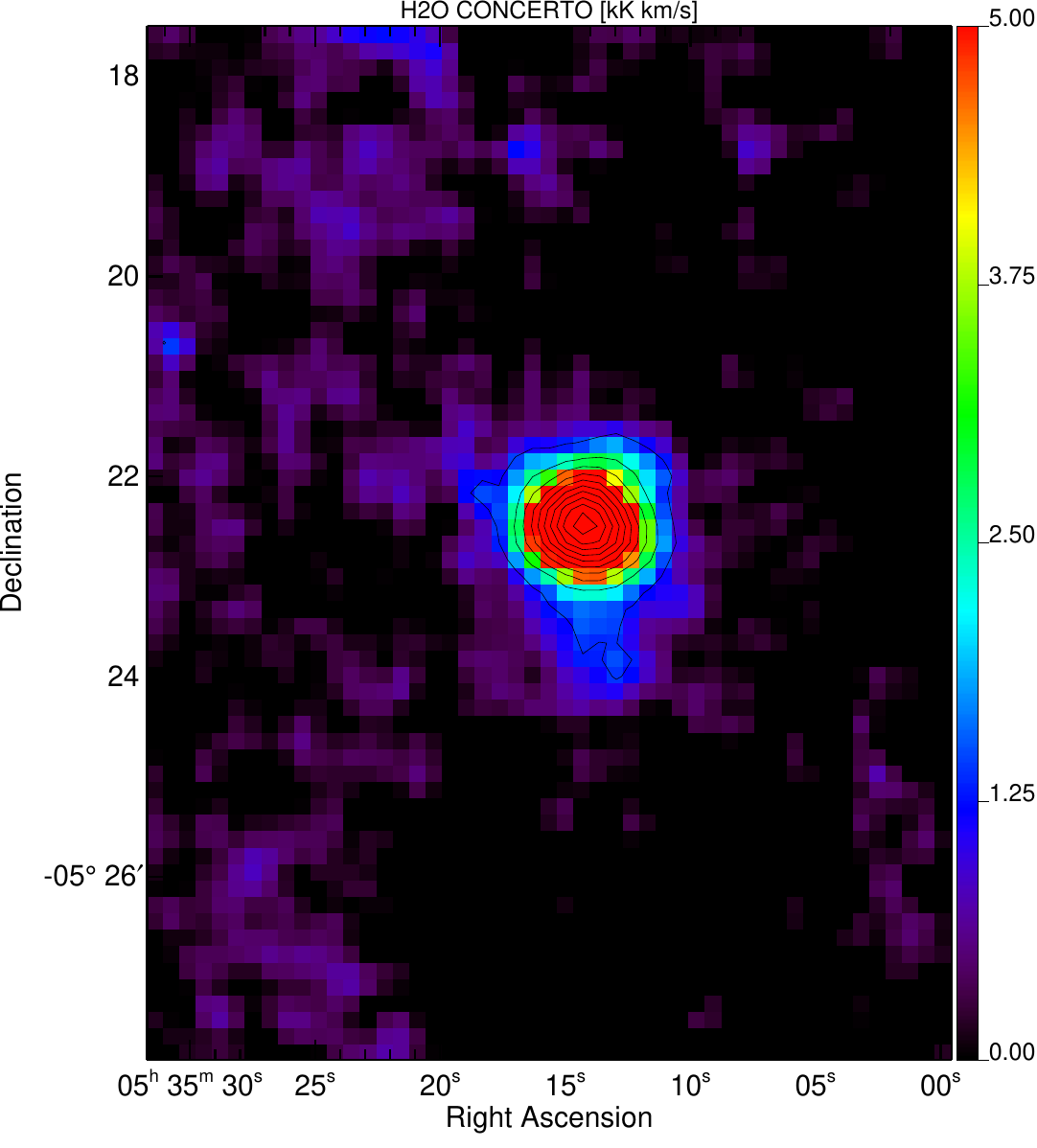}
\caption{Integrated-intensity map of the H$_2$O line in Orion, after convolution with a 20~arcsec Gaussian kernel to reach a resolution of 40~arcseconds, focussing on the Orion centre. The maximum intensity (saturated in the figure) is $14\, \mathrm{kK\,km\,s^{-1}}$. Contours indicate the S/N starting at 3 and increasing in steps of 3. 
}
\label{fig:OrionH2O}
\end{figure}

\citet{Cernicharo1994} detected an extended emission of the 183.310\,GHz $3_{13}$-$2_{20}$ para water vapour transition in Orion using the IRAM 30~m telescope. Here, we also computed and mapped the intensity of this line in our cube following exactly the same procedure as for the CO line.

Figure~\ref{fig:OrionH2O} shows a clear detection of this H$_2$O line (33~$\sigma$), localised at the very centre of Orion (Becklin-Neugebauer/Kleinmann-Low, BN/KL, 05h35m14.16s, $-05$d22'21.5") and absent from the more extended nebula.
This detection was unexpected, as ground-based telescopes are rarely transparent to that line, because with just 1~mm of water vapour in the atmosphere, the sky is effectively opaque at that frequency. The present observations, however, were made in extremely favourable weather, with 0.5~mm of pwv or below, and thus the transmission of 0.5 (Fig.~\ref{fig:bandpass}) is not negligible at 183~GHz in these conditions. One might suspect that the atmospheric water vapour emission could produce this line. However, this is unlikely because: 1)~the atmospheric emission occurs in the near-field of the telescope so is mostly common to all detectors and is removed by common mode subtraction, and 2)~the instrument's intrinsic instantaneous beam differencing cancels atmospheric emission (Eq.~\ref{eq:2beamMapSpectro}). We could also argue that the uncertainty of the bandpass, or an approximate opacity correction, produces a feature at the water frequency which, when multiplied by the Orion continuum, would simulate a line. However, that line map should be correlated to the continuum map of Orion and, in particular, the Orion Bar should show up, which is not the case.
We measure a water central intensity of $14\, \mathrm{kK\,km\,s^{-1}}$, comparable to the $\sim18\, \mathrm{kK\,km\,s^{-1}}$, reported by \citet{Cernicharo1994} (Fig. 3). We can also compare integrated emissions. 
If we assume an emission FWHM equal to 45~arcseconds, Fig.~3 of \citet{Cernicharo1994} suggest approximately $11\,\mathrm{kK\,km\,s^{-1}\,arcmin^2}$, in  satisfactory agreement (taking into account calibration uncertainties) with the $10\,\mathrm{kK\,km\,s^{-1}\,arcmin^2}$ value listed in Tab.~\ref{tab:roi}. We also note the southern extension visible in Fig.~\ref{fig:OrionH2O}, which was observed by \citet{Cernicharo1994} at velocities from 2 to 15\,km/s. Using \textit{Herschel}/HIFI, \citet{Melnick2020} have observed the Orion Nebula in the H$_2$O 1$_{10}$ – 1$_{01}$ 557~GHz transition (their Fig.~7), revealing strong emission at the BN/KL centre, the southern extension, and the Orion Bar. Unfortunately, comparing different water transition lines is difficult due to radiative-transfer issues.

\section{Discussion}\label{sec:discussion}

There is a systematic bias that comes from the relative uncertainties of the bandpass measurement. We suspect that this is what causes the wiggles seen at high frequencies (above 270~GHz), common to all panels in Fig.~\ref{fig:OrionSpectrum}. However, we do not think that this systematic bias is affecting the CO line. The best differentiating tool for that is the EW, which should be constant under bandpass bias, whereas it clearly varies in Fig.~\ref{fig:OrionEW}. Yet, the bandpass uncertainties introduce a systematic absolute calibration uncertainty to the integrated line measurement of CO.\\

For a modified blackbody with a power-law emissivity of index $\beta$ and a temperature $T_d$, it can be shown (by taking the logarithmic derivative of the Planck function) that the power-law index of the temperature brightness emission is
\begin{equation}\label{eq:EmCont2}
\alpha = \beta -\gamma\, ,
\end{equation}
with a correction index
\begin{equation}\label{eq:EmCont3}
\gamma = \frac{x\,e^x}{e^x -1} - 1\, ,
\end{equation}
which vanishes as $\frac{x}{2}\,(1-\frac{5x}{6})$ in the Rayleigh-Jeans regime, with $x=\frac{h\nu_0}{k T_{d}}$.
With intensity, rather than temperature brightness, Eq.~\ref{eq:EmCont2} would mean that the index would be $2+\beta-\gamma$. At $\nu_0=250\,\rm GHz$, the correction index is 0.33, 0.21, and 0.16 for dust temperatures of 20, 30, or 40~K. For the BN/KL region, the measurement of $\alpha = 1.10\pm0.02$ (Tab.~\ref{tab:roi}) translates into $\beta=1.31$ for 30~K dust, but this is a lower limit to the dust emissivity index, because free-free emission could contaminate the $\alpha_{250}$ value (see e.g. \citealt{Sheehan2016}). Fig.~\ref{fig:OrionAlpha} shows the strong variability of the spectral index and gives a peak at $2.0\,\pm 0.3$ for the BN/KL centre of Orion. We note that the Northern Filament globally exhibits a low emissivity index as well. 
\citet{Dicker2009} made a high-angular-resolution analysis of radio, millimetre, and sub-millimetre photometric observations, and also concluded that, even accounting for free-free emission, the dust was measured to have a low emissivity index, whether in the BN/KL (with $\beta\sim 1.2\,\pm 0.1)$ or in the main nebula. They favour a distribution of temperatures along the line of sight or a change in dust properties.
\citet{Schnee2014} measured the millimetre emissivity in Orion and also found low values for the spectral index, around 0.9, which is in some tension with the findings of \citet{Sadavoy2016}.
The Submillimetre Common-User Bolometer Array 2 (SCUBA2) instrument on the JCMT has measured a wide range of dust temperatures, spanning 10 to 50~K \citep{Salji2015}, in the so-called Orion A North integral-shaped filament. The ARTEMIS/APEX instrument and the \textit{Herschel} Gould Belt Survey have also produced temperature maps of Orion~\citep{Schuller2021}, showing 50~K temperatures in Orion Molecular Cloud OMC-1, and less than 20~K in the North Filament (OMC2 and OMC3). \citet{Nozari2025} have found anomalous dust emissivity at 3~mm by comparing NOEMA and ALMA data in some OMC-2 and OMC-3 protostellar cores, at 10-15 arcsec resolution.

The final averages of the main spectral quantities are given in Tab.~\ref{tab:roi} for the regions shown in the left panel of Fig.~\ref{fig:OrionSpectrum}. 
This table confirms that the Orion Bar is almost devoid of water vapour. Its spectrum is dominated by free-free emission because $\alpha_{200}=-3.01\pm0.10$ is lower than $-2$. The source south of BN/KL is intermediate between the BN/KL and the Bar, being a mix of dust and free-free emission. The BN/KL source is dominated by dust emission, although the power law is already flattening at 200~GHz, and it contains a strong emission of CO and H$_2$O. The Northern Filament has a rather shallow spectrum, and shows CO and H$_2$O lines. We reiterate here the argument that uncertainty in the bandpass (including the atmospheric transmission) would translate into a constant line $EW$ map. If we use the Orion Bar to estimate the bandpass-produced bias, we find $-0.4\,\%$. Taking this value as a $2\sigma$ systematic bias, it translates into a 12\ \% bias in the transmission at the H$_2$O frequency, with a constant $EW=0.1/R_{CTO}$ (where $R_{CTO}=183/3=61$ is the CONCERTO spectral resolution). Table\,\ref{tab:roi} shows much more diverse $EW$ values than the $0.3\,\%$ expected from a transmission bias. Correcting for this possible bias would amplify the H$_2$O line measurement in BN/KL. 
Conversely, possible lines at 160 and 305~GHz, seen in spectra of Fig.~\ref{fig:OrionSpectrum}, could be due to the bandpass uncertainties, because the resulting line maps match their continuum maps and because they are on the edges of the bandpass.

\citet{Serabyn1995} conducted a Fourier transform spectrometer measurement with the CSO, at 0.2~GHz resolution and a 30~arcsec beam, in the millimetre and sub-millimetre domain of Orion BN/KL. The main lines are from the CO rotational ladder. The integrated line intensity is measured at $2570\,\mathrm{K\,km\,s^{-1}}$, which is compatible with Tab.~\ref{tab:roi}, once areas are taken into account.
Other major contributors are from the HCN, SO, and SO$_2$ species. They emit more than CO, but in many widespread lines, making a pseudo-continuum across the measured CONCERTO spectrum. 
\begin{table*}[ht]
\caption{Continuum, CO, and H$_2$O results for selected regions of Orion, as defined in the left panel of Fig.~\ref{fig:OrionSpectrum}. 
}
\label{tab:roi}
\centering
\resizebox{2\columnwidth}{!}{
\begin{tiny}
\begin{tabular}{|l|l|r|r|r|r|r|r|r|r|r|}
\hline
\# & Name                    &  Size               &  $T_{250}$  & $\alpha_{250}$ &  $I_{\rm CO}$  &  $EW_{\rm CO}$ &  $T_{200}$  & $\alpha_{200}$ &  $I_{\rm H2O}$  &  $EW_{\rm H2O}$ \\ 
           &                 &      $\rm arcmin^2$   &       mK &  &      $\rm K.km.s^{-1}$ &  \%  &      mK &  &      $\rm K.km.s^{-1}$ &  \%  \\
\hline
   1 &         Main &     94.4 &     65.9 $\pm$      0.3 &     0.43 $\pm$     0.05 &     122. $\pm$       4. &    0.639 $\pm$    0.023 &     73.1 $\pm$      0.3 &    -1.81 $\pm$     0.06 &      73. $\pm$      16. &    0.290 $\pm$    0.065 \\ 
   2 &        BN/KL &      1.6 &   1124.6 $\pm$      2.1 &     1.10 $\pm$     0.02 &    1285. $\pm$      33. &    0.417 $\pm$    0.011 &   1068.7 $\pm$      2.4 &    -0.51 $\pm$     0.03 &    6356. $\pm$     123. &    1.903 $\pm$    0.037 \\ 
   3 &        South &      1.8 &    643.3 $\pm$      2.1 &     0.64 $\pm$     0.04 &     597. $\pm$      32. &    0.326 $\pm$    0.017 &    688.3 $\pm$      2.3 &    -1.19 $\pm$     0.04 &     785. $\pm$     119. &    0.346 $\pm$    0.052 \\ 
   4 &          Bar &     12.0 &     89.2 $\pm$      0.8 &    -0.39 $\pm$     0.10 &     209. $\pm$      12. &    0.759 $\pm$    0.044 &    121.6 $\pm$      0.9 &    -3.01 $\pm$     0.10 &    -175. $\pm$      45. &   -0.383 $\pm$    0.098 \\ 
   5 &  N. Filament &     34.4 &     45.8 $\pm$      0.5 &     0.23 $\pm$     0.12 &      42. $\pm$       7. &    0.312 $\pm$    0.054 &     42.5 $\pm$      0.5 &     0.18 $\pm$     0.17 &     219. $\pm$      28. &    1.746 $\pm$    0.222 \\ 
\hline
\end{tabular}
\end{tiny}
}
\tablefoot{
The measurements are all derived from the spectroscopic data cube. The quantity $T$,  is the brightness temperature of the continuum (at 250 and 200~GHz), as defined in Eq.~\ref{eq:EmCont}. The parameter $\alpha$ is the local continuum power-law index of the brightness temperature, computed between 210 and 290~GHz for the 250~GHz case, and between 165 and 220~GHz for the 200~GHz case. In both cases, the CO and H$_2$O lines are excluded from the continuum fit. The integrated line intensity, $I$,  and the effective EW, $EW$, as defined respectively by Eq.~\ref{eq:EmLine} and Eq.~\ref{eq:EW}, are given for the CO(2-1) and H$_2$O transitions. Systematic uncertainties arising from bandpass errors are not included in the table.
}
\end{table*}

\section{Conclusions and perspectives}
\label{sec:conclu}

The potential of CONCERTO spectrophotometry has been demonstrated through the creation of a map cube encompassing an area of  half a square degree, centred on the Orion Nebula at half an arcminute resolution. This map cube was obtained in just 2.4 hours of observation time. In this nebula, the millimetre spectrum is characterised by dust emission, the CO(2-1) and H$_2$O lines, and free-free emission. These components can be separated within the same dataset for the first time\footnote{Data are made public on the CONCERTO repository \url{https://mission.lam.fr/concerto}}.
The slope of the continuum is determined independently of the lines and consequently represents the millimetre emissivity of dust once free-free emission has been subtracted and the dust temperature is sufficiently large or moderately well known. Comparison with external datasets shows that the CONCERTO data are calibrated at the 15\,\% level.

This is the first time that KIDs have detected line emissions in spectroscopic imaging mode (DESHIMA; \citealt{Endo2019}) is a KID-based spectrograph with only one beam on the sky). The CO line mapping was achieved in a small amount of observing time relative to heterodyne one-beam instruments, albeit without velocity resolution. The H$_2$O line mapping, which shows a strong concentration in Orion BN/KL, is a serendipitous addition to this study. Limitations still remain in sensitivity (severe cuts in the data) and in bandpass knowledge, which will be addressed in future publications.

The CONCERTO observations of the water vapour line in Orion are in agreement with \citet{Cernicharo1994} heterodyne measurements. Some water vapour may even be detected in the Northern Filament. The main explanation for the strong presence of water vapour in the BN/KL object is the sublimation of ice water mantles from heated or shocked grains~\citep{vanDishoeck2013}. This makes water vapour the main coolant of the warm molecular clouds~\citep{Cernicharo2005}. Observations of Orion sub-millimetre lines of water vapour have been reported by \citet{Tauber1996}, \citet{Harwit1998}, \citet{Snell2000}, \citet{Olofsson2003}, and \citet{Goicoechea2015}.

The Galactic Centre, mapped over 2 square degrees, was observed in the same ESO programme and will be reported in another publication, along with more elaborate component spectral separation methods. Indeed, Orion proved to be a challenging region to analyse in terms of dust emissivity because the free-free component is relatively important.

\begin{acknowledgements}
        Besides the authors, the technicians and engineers more involved in the CONCERTO experimental setup development have been Maurice Grollier, Olivier Exshaw, Anne Gerardin, Gilles Pont, Guillaume Donnier-Valentin, Philippe Jeantet, Mathilde Heigeas, Christophe Vescovi, and Marc Marton. We acknowledge the crucial contributions of the whole Cryogenics and Electronics groups at Institut Néel and LPSC. We acknowledge the contribution of Hamdi Mani, Chris Groppi, and Philip Mauskopf (from the School of Earth and Space Exploration and Department of Physics, Arizona State University) to the cold electronics. The KID arrays of CONCERTO have been produced at the PTA Grenoble microfabrication facility. We warmly thank the support from the APEX staff for their help in CONCERTO pre-installations and design. The flexible pipes, in particular, have been routed under the competent coordination of Jorge Santana and Marcelo Navarro. We acknowledge support from the European Research Council (ERC) under the European Union’s Horizon 2020 research and innovation programme (project CONCERTO, grant agreement No 788212), from the Excellence Initiative of Aix-Marseille University-A*Midex, a French "Investissements d’Avenir" programme, from the LabEx FOCUS ANR-11-LABX-0013, and from the ECOS-ANID French and Chilian cooperation program.  DQ acknowledges support from the National Agency for Research and Development (ANID)/Scholarship Program/Doctorado Nacional/2021-21212222 and ECOS-ANID N$^{\rm{o}}$ECOS220016. This work has also been supported by the GIS KIDs. We are grateful to our administrative staff in Grenoble and Marseille, in particular Patricia Poirier, Mathilde Berard, Lilia Todorov and Valérie Favre, and the Protisvalor team. We acknowledge the crucial help of the Institut Néel and MCBT Heads (Etienne Bustarret, Klaus Hasselbach, Thierry Fournier, Laurence Magaud) during the COVID-19 restriction period. MA acknowledges support from FONDECYTgrant 1211951 and ANID BASAL project FB210003. E.I. gratefully acknowledges financial support from ANID - MILENIO - NCN2024\_112 and ANID FONDECYT Regular 1221846. \\

The present results are based on observations collected at the European Organisation for Astronomical Research in the Southern Hemisphere under ESO programme 110.23NK. 

This research has made use of the SIMBAD database, operated at CDS, Strasbourg, France.

This work began while one of us (FXD) was at the Kavli Institute for Cosmology (Cambridge) on a 6-month grant from the French Embassy in the UK and Churchill College.\\
We would like to thank the anonymous referee for their clarification requests.\\

We would like to thank the IRAM staff for their support during the campaigns. The NIKA2 dilution cryostat has been designed and built at the Institut Néel. In particular, we acknowledge the crucial contribution of the Cryogenics Group, and in particular Gregory Garde, Henri Rodenas, JeanPaul Leggeri, Philippe Camus. This work has been partially funded by the Foundation Nanoscience Grenoble and the LabEx FOCUS ANR-11-LABX-0013. This work is supported by the French National Research Agency under the contracts “MKIDS”, “NIKA” and ANR-15-CE31-0017 and in the framework of the “Investissements d’avenir” program (ANR-15-IDEX-02). 
\end{acknowledgements}

\bibliographystyle{aa}
\bibliography{aa55320-25}

\end{document}